\documentclass[12pt]{article}
\usepackage{a4wide}
\usepackage[centertags]{amsmath}
\usepackage{amssymb}
\usepackage[sort&compress,numbers]{natbib}
\usepackage{ifpdf}
\usepackage{setspace}
\usepackage{amsfonts}
\usepackage{color}
\usepackage{threeparttable}
\usepackage{cancel}
\usepackage{tikz}
\usetikzlibrary{decorations.pathmorphing}
\usetikzlibrary{decorations.markings}
\usetikzlibrary{arrows}
\usepackage{feynmf}
\usepackage[compat=1.1.0]{tikz-feynman}
\tikzfeynmanset{compat=1.1.0}
\usepackage{verbatim}
\usepackage{xcolor}
\usepackage{ulem}

\usepackage{lscape}

\usepackage{blindtext, rotating}

\usepackage{yfonts}

\usetikzlibrary{decorations.markings}
\ifpdf
\usepackage[colorlinks=true,
linkcolor=black,
citecolor=black,
urlcolor=blue,
filecolor=blue,
pdfstartview=FitV,
pdftitle={},
pdfauthor={},
pdfsubject={Hydrodynamics},
pdfkeywords={hydrodynamics, AdS/CFT},
pdfpagemode=None,
bookmarksopen=true
]{hyperref}
\else
\usepackage[hypertex]{hyperref}
\fi
\usepackage{rotating}
\usepackage{rotating}
\makeatletter
\@addtoreset{equation}{section}

\newcommand{\qn}{{\mathfrak{q}}}
\newcommand{\wn}{{\mathfrak{w}}}

\makeatletter
\renewcommand\section{\@startsection {section}{1}{\z@}%
	{-3.5ex \@plus -1ex \@minus -.2ex}
	{2.3ex \@plus.2ex}%
	{\normalfont\large\bfseries}}
\renewcommand\subsection{\@startsection{subsection}{2}{\z@}%
	{-3.25ex\@plus -1ex \@minus -.2ex}%
	{1.5ex \@plus .2ex}%
	{\normalfont\bfseries}}

\def\sec#1{\S \;\ref{#1}}

\title{{On the correlation functions in stable first-order relativistic hydrodynamics}}
\author{Navid Abbasi$^a$\footnote{abbasi@lzu.edu.cn}, \,\,\,Ali Davody$^a$\footnote{adavody@soundhound.com}, \,\,\,Sara Tahery$^{b}$\footnote{saratahery@htu.edu.cn} \\
	\small{\emph{}}\\
	\small{$^{a}$School of Nuclear Science and Technology, Lanzhou University,}\\
	\small{
		222 South Tianshui Road, Lanzhou 730000, China } \\
	\small{$^{b}$SoundHound AI, Inc,}\\
	\small{
		5400 Betsy Ross Drive
		Santa Clara, CA 95054, USA } \\
	\small{\emph{}	{ 
			$^{c}$Institute of Particle and Nuclear Physics,  
		}} \\
	\small{
		Henan Normal University,  Xinxiang 453007, China }\\
}
\begin{document}
	
	\setlength{\baselineskip}{16pt}
	\begin{titlepage}
		\maketitle
		
		\vspace{-36pt}
		
		\begin{abstract}
			First-order relativistic conformal hydrodynamics in a general (hydrodynamic) frame is characterized by a shear viscosity coefficient and two UV-regulator parameters. 
            Within a certain range of these parameters, the equilibrium is stable and propagation is causal. In this work we study the correlation functions of fluctuations in this theory. We first compute hydrodynamic correlation functions in the linear response regime. Then we use the linear response results to explore the analytical structure of response functions beyond the linear response. A method is developed to numerically calculate the branch cut structure from the well-known Landau equations. We apply our method to the shear channel and find the branch cuts of a certain response function, without computing the response function itself. We then solve the Landau equations analytically and find the threshold singularities of the same response function. Using these results, we achieve the leading singularity in momentum space, by which, we find the long-time tail of the correlation function. The results turn out to be in complete agreement with the loop calculations in effective field theory. 
			
		\end{abstract}
		\thispagestyle{empty}
		\setcounter{page}{0}
	\end{titlepage}

	\renewcommand{\baselinestretch}{1}  
	\tableofcontents
	\renewcommand{\baselinestretch}{1.2}  
	\section{Introduction}
	It is well known that relativistic hydrodynamics beyond zeroth order in derivatives can be constructed in various ways,  depending on how the fluid velocity and temperature are defined when the system is not in equilibrium	\cite{Kovtun:2012rj}. Any particular choice of definition for these out of equilibrium variables is called a choice of ``frame''\footnote{It does not have to be confused with the concept of Lorentz frames.}. In the famous Eckart 	\cite{Eckart:1940te}	and Landau-Lifshitz  	\cite{Landau:Fluid}  frames, hydrodynamics exhibits some non-physical features: the thermal equilibrium is unstable 	\cite{Hiscock:1985zz} and the theory supports the propagation of superluminal modes 	\cite{Hiscock:1987zz}.
	\vspace{2mm}
	
	Problems of instability and acausality  are associated with modes outside the range of validity of hydrodynamics, i.e., beyond the long wavelength low frequency limit. In other words, the problem is due to the UV modes.
	To remedy this, one way is to introduce extra dynamical degrees of freedom into hydrodynamics, thereby modifying the behavior of the theory in the UV. This is actually the idea of the Israel-Stewart theory \cite{Israel,Israel:2}. Another way to deal with this problem is to include the second order derivative corrections \cite{Baier:2007ix,Luzum:2008cw,Denicol:2012cn} \footnote{In practice, causal relativistic hydrodynamics can describe the evolution of hot QCD matter formed in heavy ion collisions (see \cite{Florkowski:2017olj,Shen:2020mgh} for a review).}.
	A more systematic way of dealing with this problem is to find certain out of equilibrium definitions for hydrodynamic variables; something other than those used  by Eckart or Landau-Lifshitz.  This program was touched on by 		\cite{Van:2011yn,Freistu:2011yn}, 
	and built in  \cite{Bemfica:2017wps} and \cite{Kovtun:2019hdm}. These theories (\cite{Bemfica:2017wps,Kovtun:2019hdm}) are sometimes referred to as BDNK in the literature \footnote{ The name stands for Bemfica-Disconzi-Noronha-Kovtun. The idea of relativistic hydrodynamics in general frames  \cite{Kovtun:2019hdm,Bemfica:2020zjp,Bemfica:2019knx} has been extended in several directions, by inclusion of charge, magnetic field, and spin \cite{Armas:2022wvb,Biswas:2022cla,Hoult:2021gnb,Hoult:2020eho, Mitra:2021ubx,Shokri:2020cxa,Shokri:2023rpp}, and also has been compared with other causal and stable frameworks \cite{Daher:2022wzf,Das:2020fnr,Das:2020gtq,Freistuhler:2021lla}. See also \cite{Dore:2021xqq}; a gauge theory perspective on stable theories of hydrodynamics. }.
		\vspace{2mm}
		
In this work, we study correlation functions in BDNK theory. We first proceed to calculate the correlation function in the linear response regime. It is discussed in \sec{linear_response}. We perform calculations separately for the sound and shear channels. The pole structure of the correlation function is found to be completely consistent with the linear excitation spectrum found in \cite{Kovtun:2019hdm}. Furthermore, our calculations reveal an important feature of BDNK theory. We find that the special form of the BDNK equation causes the energy density correlator to develop a range of negative values. This can happen at any fixed value of momentum when we take a small frequency limit. However, this does not happen when we take the small momentum limit at a fixed frequency.
 		\vspace{2mm}
 		
			Apart from  some numerical works (including \cite{Pandya:2021ief,Bantilan:2022ech}), little is known about BDNK theory in the `nonlinear' regime. 	The second part of this work is devoted to study the nonlinear fluctuations in BDNK theory. The main question is how the interactions between hydrodynamic modes affect the late-time behavior of the correlation functions. In conventional hydrodynamics, this question has a well-known answer. Nonlinear fluctuations lead to longtime tails with fractional powers in the real space correlation function \cite{Ernst,Kovtun:2003vj}. This is actually due to the nonlinearity that causes branch point singularities in the correlation function \cite{Kovtun:2012rj,Chen-Lin:2018kfl,Martinez:2018wia}. In this work, we want to explore what happens to the longtime tails in BDNK theory.
					\vspace{2mm}
					
			   In a systematic approach, this can be explored within the framework of the effective field theory of hydrodynamics. However, we choose a different route here. 
	We use the fact that the correlation function of certain operators in the nonlinear regime factorize if the distribution of fluctuations is taken to be Gaussian \cite{Kovtun:2003vj}.  In quantum field theory (QFT), the factorized correlation function is in the form of a Feynman integral, representing a simple loop diagram with two lines, i.e., a bubble diagram.
	We then follow the famous Landau method to find the singularities of the Feynman integral. Instead of computing the integral explicitly, Landau proposed a set of equations, the so-called ``\textit{Landau equations}", which are solved to specify the threshold singularities of the Feynman integral.
		\vspace{2mm}
		
	Keeping in mind the ideas discussed in the previous paragraph, we explore the analytic structure of the shear stress response function in \sec{non_linear}. For some cases in effective field theory (EFT) of hydrodynamic interactions, knowing the analytic structure of the hydrodynamic response functions, including the threshold singularities as well as the branch cut structure, is sufficient to find decay rates or cross sections without performing Feynman integration formal way.
	 This is discussed at the beginning of \sec{reason}.  In \sec{branch_cut}, we develop a method to numerically find the location of branch cuts.  To specify threshold singularities, we derive the associated Landau equations and solve them analytically in \sec{threshold}. We show that our results are in complete agreement with recent EFT results obtained from explicit loop calculations.  
		\vspace{2mm}
		
	  Based on these results, we will calculate the longtime tail of shear stress correlation function in \sec{long_time}. We find that within the validity range of BDNK theory, the long-time tail of the shear stress correlation function in BDNK theory is consistent with the long-time tail of the shear stress correlation function in conventional hydrodynamics. 
	\vspace{2mm}

	Finally, in \sec{conclusion} we end with a review of the results, mentioning possible applications and discussing some followup directions.
	\\\\
	\textbf{Note:} Throughout this paper we will refer to ``Landau'' in two contexts: 
	\begin{enumerate}
		\item ``Landau-Lifshitz frame'' (or  Landau-Lifshitz hydrodynamics), which refers to some specific way of defining the fluid four-velocity beyond zeroth order in the derivative expansion. 
		\item  ``Landau equations/conditions'' refer to those equations that allow finding the singularity of the Feynman integral without explicitly performing the integration.
	\end{enumerate}
	It should be emphasized that the above two items have nothing to do with each other.
	\section{First-order stable conformal hydrodynamics}
	\label{stabel_hydro}
	In an uncharged conformal fluid, the constitutive relations in the most general frame are determined by four dimensionless numbers $\bar{p}$, $\bar{\pi}$, $\bar{\theta}$ and $\bar{\eta}$ \cite{Kovtun:2019hdm}. Specializing to $D=4$ dimensions, one writes
	\begin{equation}\label{T_mu_nu_Covariant_Kovtun}
		\begin{split}
			T^{\mu\nu}=&\,T^3\left(\bar{p}\,T+\,\bar{\pi}\frac{u^{\lambda}\partial_{\lambda}T}{T}+\,\bar{\pi}\frac{\partial_{\lambda} u^{\lambda}}{3}\right)(\eta^{\mu\nu}+\,4\,u^{\mu}\,u^{\nu})\\
			&+\bar{\theta}\,T^3\left[\left(u^{\lambda}\partial_{\lambda}u^{\mu}+\,\frac{P^{\mu\lambda}\partial_{\lambda}T}{T}\right)u^{\nu}+\,(\mu \leftrightarrow\nu)\right]-\,\bar{\eta}\,T^3\,\sigma^{\mu\nu}\,.
		\end{split}
	\end{equation}
	Note that $p=\bar{p}T^4$ is the equilibrium pressure and $\eta=\bar{\eta}T^3$ is the shear viscosity, both depending of the microscopic of the fluid. Two new transport coefficients, $\pi=\bar{\pi}T^3$ and $\theta=\bar{\theta}T^3$ are responsible for regulating the theory in the UV limit. 
		\vspace{2mm}
		
	The stability and causality of the first order hydrodynamics has been discussed extensively in the literature, over the past few years \cite{Bemfica:2017wps,Kovtun:2019hdm,Armas:2022wvb,Bantilan:2022ech,Bemfica:2019knx,Biswas:2022cla,Daher:2022wzf,Das:2020fnr,Das:2020gtq,	Freistuhler:2021lla,Hoult:2020eho,Hoult:2021gnb,Mitra:2021ubx,Pandya:2021ief}. In the case of conformal fluid in $D=4$ dimensions, these conditions are given by \cite{Bemfica:2017wps}
	\begin{equation}\label{stability}
		1-\frac{3\bar{\eta}}{\bar{\theta}}-\frac{\bar{\eta}}{\bar{\pi}}>0\,,\,\,\,\,\,\bar{\pi}>4 \bar{\eta}\,.
	\end{equation}
	To satisfy \eqref{stability}, it is sufficient to take $\bar{\theta}>4 \bar{\eta}$ and $\bar{\pi}>4\bar{\eta}$ \cite{Kovtun:2019hdm}.
	These conditions ensure that in any equilibrium state specified by $u^{\mu}=\gamma(1,\textbf{v}_0)$ and $T=T_0$,  linearised evolution of \eqref{T_mu_nu_Covariant_Kovtun} is always dissipative and in the light-cone.
	Let us denote that the relativistic hydrodynamics in a general frame, i.e., \eqref{T_mu_nu_Covariant_Kovtun},  together with the conditions \eqref{stability} is usually referred to as the ``stable first-order relativistic hydrodynamics'' theory.
	\section{Correlation functions in the linear regime}
	\label{linear_response}
	As mentioned in the Introduction, the focus of this work is on correlation functions. In this section, we explore the correlation functions in the linear regime. To this end, we take $T^{\mu\nu}$ from \eqref{T_mu_nu_Covariant_Kovtun} and linearize the hydrodynamic equations, namely $\partial_{\mu}T^{\mu\nu}$=0, around the equilibrium state specified by $u^{\mu}=\gamma(1,\textbf{v}_0)$ and $T=\text{constant}$. The linearised equations then govern the coupled linear evolution of  $\delta\textbf{v}$ and $\delta T$, as follows.
	\vspace{2mm}
	
	In conventional hydrodynamic treatments, taking $\textbf{v}_0\ne0$ is especially important to make  the instabilities of the equilibrium manifest \cite{Hiscock:1985zz}. However, in this work, we restrict our study to the linearised equations when $\textbf{v}_0=0$. 
	\vspace{2mm}

	The conservation of energy, $\partial_{\mu}T^{\mu t}=0$ gives
	\begin{equation}\label{EoM_t}
		\underline{	\partial_t \frac{\delta T}{T}+\frac{1}{3}\boldsymbol{\nabla}\cdot\delta\textbf{v}}+\frac{1}{12\bar{p}T}\left[\bar{\theta}\,\boldsymbol{\nabla}\cdot \left(\underline{\underline{\partial_t\delta\textbf{v}+\boldsymbol{\nabla} \frac{\delta T}{T}}}\right)+3 \bar{\pi}\,\partial_t\left(\underline{	\partial_t \frac{\delta T}{T}+\frac{1}{3}\boldsymbol{\nabla}\cdot\delta\textbf{v}}\right)\right]=\,0
	\end{equation}
	The equations of momentum conservation, $\partial_{\mu}T^{\mu i}=0$, give
	\begin{equation}\label{EoM_z}
		\underline{\underline{\partial_t\delta\textbf{v}+\boldsymbol{\nabla} \frac{\delta T}{T}}}+\frac{1}{4 \bar{p}T}\left[\bar{\theta}\partial_t \left(\underline{\underline{\partial_t\delta\textbf{v}+\boldsymbol{\nabla} \frac{\delta T}{T}}}\right)+\bar{\pi}\boldsymbol{\nabla}\left(\underline{\partial_t \frac{\delta T}{T}+\frac{1}{3}\boldsymbol{\nabla}\cdot\delta\textbf{v}}\right)-\frac{4}{3}\bar{\eta}\boldsymbol{\nabla} ^2\delta\textbf{v}\right]=\,0
	\end{equation}
	The first two terms in each equation come from $T^{\mu\nu}$ of order zero, while the terms in brackets are associated with $T^{\mu\nu}$ of order one. It is well known, and evident in the above equations, that the stable theory of hydrodynamics is constructed in such a way that the zeroth order equations emerge ``with new transport coefficients $\bar{\theta}$ and $\bar{ \pi }$'' at the first order. If one proceeds to solve the equations perturbatively, that is, order by order in the derivative expansion, the above two coefficients will be washed out at the first order simply because the zeroth order equations are used. However, this differs from the concept of ``stable relativistic hydrodynamics", where the equations must be solved non-perturbatively.
	\subsection{How to calculate correlation functions?}
In the hydrodynamics literature, response functions are computed within the framework of the linear response theory \cite{Martin}\cite{Kovtun:2012rj}. 
To find a general response function such as $G_{\mathcal{R}}(t,\textbf{k})$ we need to solve a differential equation of order one.  This is because the conventional relativistic hydrodynamic equations in an uncharged system, the Landau-Lifshitz equations, are `\textit{parabolic}'. The result can then be used to find the correlation function via the fluctuation-dissipation theorem. 
	\vspace{2mm}

However, in the ``general frame'' formalism \cite{Kovtun:2019hdm}, the hydrodynamic equations are `\textit{hyperbolic}'. Computing the response function in this case requires an extension of the standard method developed in \cite{Martin,Kovtun:2012rj} to second-order time-differential equations. Here we choose a different rout; instead of computing the response function, we choose to compute the correlation function directly.
To this end, we extend the Landau-Lifshitz method of computing the correlation function \cite{Landau_1,Landau_2} to relativistic systems.  We can then compute various correlation functions. With them, one can also use fluctuation-dissipation theorem to find the response functions. In this work, however, our focus is on computing correlation functions. This is what we will do in the next two subsections.
	\subsection{Correlation functions in longitudinal channel }
	An important aspect of the ``stable relativistic hydrodynamics" is that it allows us to treat fluid velocity and temperature as the hydrodynamic variables, employing the same concepts we know in thermodynamics; and similar to what is done in non-relativistic fluid dynamics. 
	Then this suggests that in addition to the correlation functions of conserved charges (or currents), the correlation function of hydrodynamic variables, such as $\langle v_{i}(t_1,\textbf{x}_1)v_j(t_2,\textbf{x}_2) \rangle$, may also have sensible description.
	It should be emphasized that our main physical arguments in this paper will be made based on the correlation function of conserved densities. However, as we will explain below, each of the latter correlators is a linear combination of former ones.
		\vspace{2mm}
	
		Keeping in mind  the above discussion, we first compute the correlation function of the hydrodynamic variables.
	Considering $\delta \phi_{a}=(\frac{\delta T}{T}, \delta v_i)\equiv(\frac{\delta T}{T}, \delta v^{\perp}, \delta v^{\parallel})$, the correlation function of $\phi_a$ and $\phi_b$ is defined as
	\begin{equation}\label{Fourier}
		\big\langle\delta \phi_{a}\delta \phi_{b}\big\rangle_{\omega\textbf{k}}=\int_{-\infty}^{+\infty} dt\int d^3 \text{x} \,e^{i \omega t} e^{-i \textbf{k}\cdot \textbf{x}} \langle \delta \phi_{a}(t, \textbf{x}) \delta \phi_{b}(0, \textbf{0})\rangle \,.
	\end{equation}
	We also find it useful to define  dimensionless quantities as 
	\begin{equation}\label{re_scale}
		\wn=\frac{\bar{\eta}}{\bar{p}}\,\frac{\omega}{T}\,,\,\,\,\,\,\,\qn=\frac{\bar{\eta}}{\bar{p}}\,\frac{\textbf{k}}{T}\,,\,\,\,\,\,\,\pi_{\eta}=\frac{\bar{\pi}}{\bar{\eta}}\,,\,\,\,\,\,\,\theta_{\eta}=\frac{\bar{\theta}}{\bar{\eta}}\,.
	\end{equation}
	As mentioned before, we extend the method developed in \cite{Landau_1,Landau_2} to compute various correlation functions $\big\langle\delta \phi_{a}\delta \phi_{b}\big\rangle_{\wn\qn}$ relativistic systems. Details can be found in  Appendix \ref{hydro_fluc_App}. 
	Interestingly we find that \footnote{Similar to \eqref{tv_vt}, one can easily compute $\big\langle\frac{\delta T}{T}\frac{\delta T}{T}\big\rangle_{\wn\qn}$ and $\big\langle\delta v_z\delta v_z\big\rangle_{\wn\qn}$. More details can be found in Appendix \ref{hydro_fluc_App}. Note that in this Appendix we take $\delta v^{\parallel}\equiv \delta v_z$. }
	\begin{equation}\label{tv_vt}
		\begin{split}
			\big\langle\delta v^{\parallel}\frac{\delta T}{T}\big\rangle_{\wn\qn}=&\frac{3\bar{\eta}}{4\bar{p}^2T^4}\frac{1}{\mathcal{D}(\wn,\qn)} \,iq\,(i \,\boxed{\pi_{\eta}}\, \wn-4)\,\big(4-i (\theta_{\eta}+\pi_{\eta})\wn\big) \\
			&+\frac{3\bar{\eta}}{4\bar{p}^2T^4}\frac{1}{\mathcal{D}^*(\wn,\qn)} \,(-iq)\,(-i\,\boxed{\theta_{\eta}}\, \wn-4)\,\big(4+i (\theta_{\eta}+\pi_{\eta})\wn\big)\,,\\
			\big\langle\frac{\delta T}{T}\delta v^{\parallel}\big\rangle_{\wn\qn}=&\frac{3\bar{\eta}}{4\bar{p}^2T^4}\frac{1}{\mathcal{D}(\wn,\qn)} \,iq\,(i\,\boxed{\theta_{\eta}}\, \wn-4)\,\big(4-i (\theta_{\eta}+\pi_{\eta})\wn\big) \\
			&+\frac{3\bar{\eta}}{4\bar{p}^2T^4}\frac{1}{\mathcal{D}^*(\wn,\qn)} \,(-iq)\,(-i\,\boxed{\pi_{\eta}}\, \wn-4)\,\big(4+i (\theta_{\eta}+\pi_{\eta})\wn\big) ]\,.
		\end{split}
	\end{equation}
	Regardless of whether the stability and causality conditions are satisfied, it is clear from \eqref{tv_vt} that for general values of $\theta_{\eta}$ and $\pi_{\eta}$:
	\begin{equation}\label{inequality}
		\big\langle\delta v^{\parallel}\frac{\delta T}{T}\big\rangle_{\wn\qn}\ne\big\langle\frac{\delta T}{T}\delta v^{\parallel}\big\rangle_{\wn\qn}\,.
	\end{equation}
	If these correlators are treated as microscopic correlation functions, then ``\textit{time-reversal invariance}'' requires that they must be equal. The latter leads to
	\begin{equation}\label{constarint}
		\pi_{\eta}=\,\theta_{\eta}\,.
	\end{equation}
	This constraint simply tells us that the time-reversal-invariant stable frames occupy a zero-measure subspace in the entire space of stable frames identified by \eqref{stability}, i.e., like a curve lying on the plane. However, this is not the case. If we implement derivative expansion, the above two correlation functions are equal to first order in the derivative, independent of the values of $\pi_{\eta}$ and $\theta_{\eta}$. In other words, the inequality \eqref{inequality} occurs outside the domain where only first-order dissipative corrections are sufficient to describe the dynamics \footnote{In order to understand how microscopic time-reversal invariance prevents \eqref{inequality} at second order in derivative, we need to have at least recourse to the second-order stable and causal theory of relativistic hydrodynamics, which is indeed beyond the scope of the present work. One type of such theories has been recently proposed in \cite{Noronha:2021syv}. It is shown that the first order truncation of this theory reduces to the BDNK theory. }.  
		\vspace{2mm}
	
\begin{figure}[tb]
		\centering
		\centering
		\includegraphics[width=0.48\textwidth]{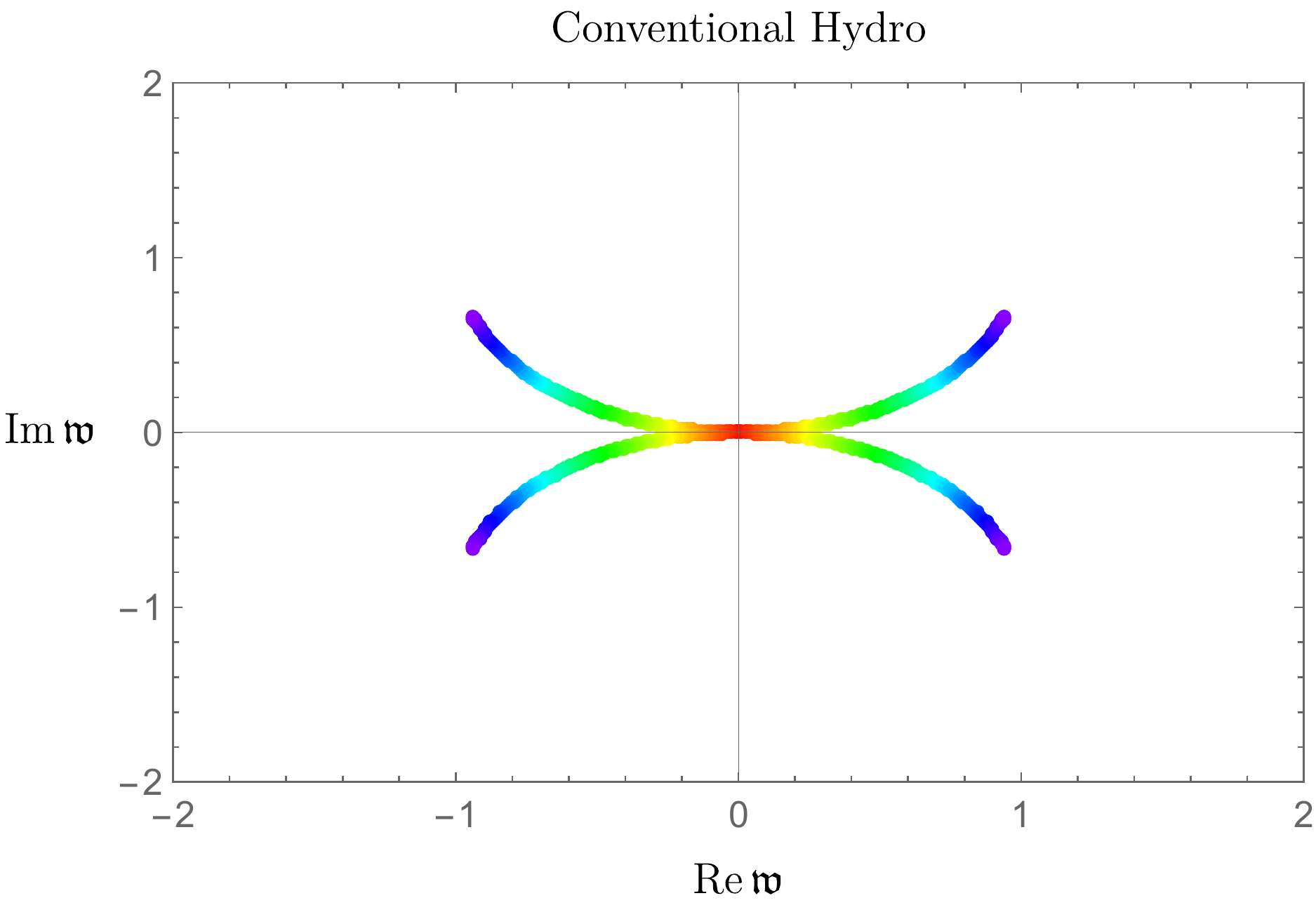}\,\,\,\,\,\includegraphics[width=0.48\textwidth]{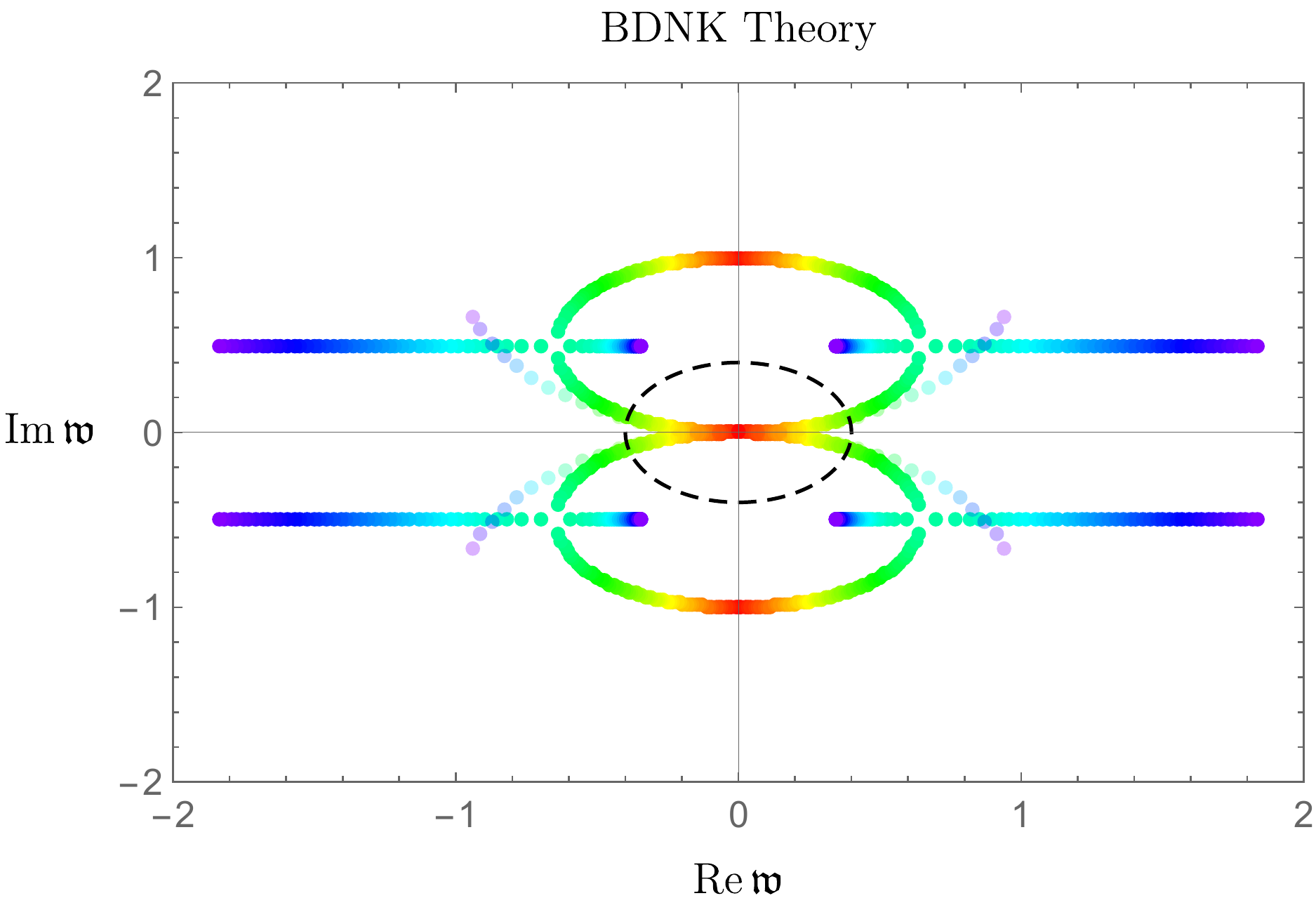}\,
		\caption{Correlation functions poles comparison between conventional relativistic hydrodynamics and BDNK theory in sound channel. Each colorful trajectory starting in red and ending in purple illustrates change in one single mode when $\qn$ varies from $0$ to $2$. Lower half plane modes correspond to poles of retarded Green's functions in sound channel, while the upper half plane modes correspond to the poles of advanced Green's functions. Conventional hydro Green's functions have only two sound modes in lower half plane while BDNK theory Green's functions have two extra modes associated with the inclusion of two parameters $\theta$ and $\pi$.  Note that to make comparison easier, we show a low-opacity version of the left panel plot on the right plot. The dashed circle shows the momentum range where the BDNK spectrum is simply consistent with the conventional relativistic hydrodynamics spectrum, i.e. $\qn\lesssim0.5$ when $\theta_{\eta}=4$.}
		\label{pole_sound}
	\end{figure}

	Now let us move on to calculating the energy density correlation function, namely  $G_{T_{tt}T_{tt}}(\wn,\qn)=\langle T^{tt}T^{tt} \rangle_{\wn \qn}$.  Using \eqref{T_mu_nu_Covariant_Kovtun}, one writes
	\begin{equation}\label{G_first}
		\begin{split}
			G_{T_{tt}T_{tt}}(\wn,\qn)=\bar{p}^2T^8&\bigg[9(\pi_{\eta}^2\wn^2+16)\big\langle\frac{\delta T}{T}\frac{\delta T}{T}\big\rangle_{\wn\qn}+\pi_{\eta}^2\, q_i q_j \big\langle\delta v_i\delta v_j\big\rangle_{\wn\qn}\\
			\,\,\,\,\,\,\,\,\,\,\,&\,\,\,\,\,-3\pi_{\eta}q_j\bigg(\big(\pi_{\eta}\wn+4i\big)\big\langle\frac{\delta T}{T}\delta v_j\big\rangle_{\wn\qn}+\big(\pi_{\eta}\wn-4i\big)\big\langle\delta v_j\frac{\delta T}{T}\big\rangle_{\wn\qn}\bigg)\bigg]\,.
		\end{split}
	\end{equation}
	Then by using the corresponding correlation functions of the hydrodynamic variables (Appendix \ref{hydro_fluc_App}), we find:
	\begin{equation}\label{G_TT_TT}
		\begin{split}
			G_{T_{tt},T_{tt}}=\frac{-6\bar{\eta}T^4 \qn^2}{|\mathcal{D}_L(\wn, \qn)|^2} &\bigg[+\left(\pi _{\eta }-4\right) \pi _{\eta }^2 \theta _{\eta }^2 \,\qn^6+96 \left(\pi _{\eta }^2-4 \pi _{\eta }+8\right) \theta _{\eta } \,\qn^4+2304 \left(\pi _{\eta }-4\right) \qn^2 \\
			&\,\,\,\,\,\,+3 \wn^2  \bigg(-2304 \pi _{\eta }+\pi _{\eta } \theta _{\eta } \left(4 \theta _{\eta }^2-3 \left(\pi _{\eta }-4\right) \pi _{\eta } \theta _{\eta }-8 \left(\pi _{\eta }-2\right) \pi _{\eta }\right) \qn^4\\
			&\,\,\,\,\,\,\,\,\,\,\,\,\,\,\,\,\,\,\,\,\,\,\,\,\,\,+48 \left(-2 \left(\pi _{\eta }-2\right) \pi _{\eta } \theta _{\eta }+\left(\pi _{\eta }-4\right) \theta _{\eta }^2+\left(\pi _{\eta }-8\right) \pi _{\eta }^2\right) \qn^2\bigg)\\
			\,+	9 \wn^4 \big(4 \pi _{\eta }^3 &\theta _{\eta } \qn^2+\pi _{\eta } \theta _{\eta }^2 \left(\pi _{\eta } \left(3 \pi _{\eta }-4\right) \qn^2-48\right)-4 \pi _{\eta }^3 \left(\pi _{\eta } \qn^2+12\right)\big)-27 \wn^6\,\pi _{\eta }^3  \theta _{\eta }^2\bigg] \,.
		\end{split}
	\end{equation}
	with 	
	\begin{equation}\label{D_L}
		\begin{split}
			\mathcal{D}_L(\wn, \qn)=&\,9 \pi _{\eta }  \,\theta _{\eta }\,\wn^4+36i  \left(\theta _{\eta }+\pi _{\eta }\right)\, \wn^3-6 \left(\pi _{\eta } \left(\theta _{\eta }+2\right) \qn^2+24\right)\wn^2\\
			&-12 i   \left(\theta _{\eta }+\pi _{\eta }+4\right) \qn^2\wn+ \left(\left(\pi _{\eta }-4\right) \theta _{\eta } \qn^2+48\right)\qn^2\,.
		\end{split}
	\end{equation}
	\begin{figure}[tb]
		\centering
		\centering
		\includegraphics[width=0.6\textwidth]{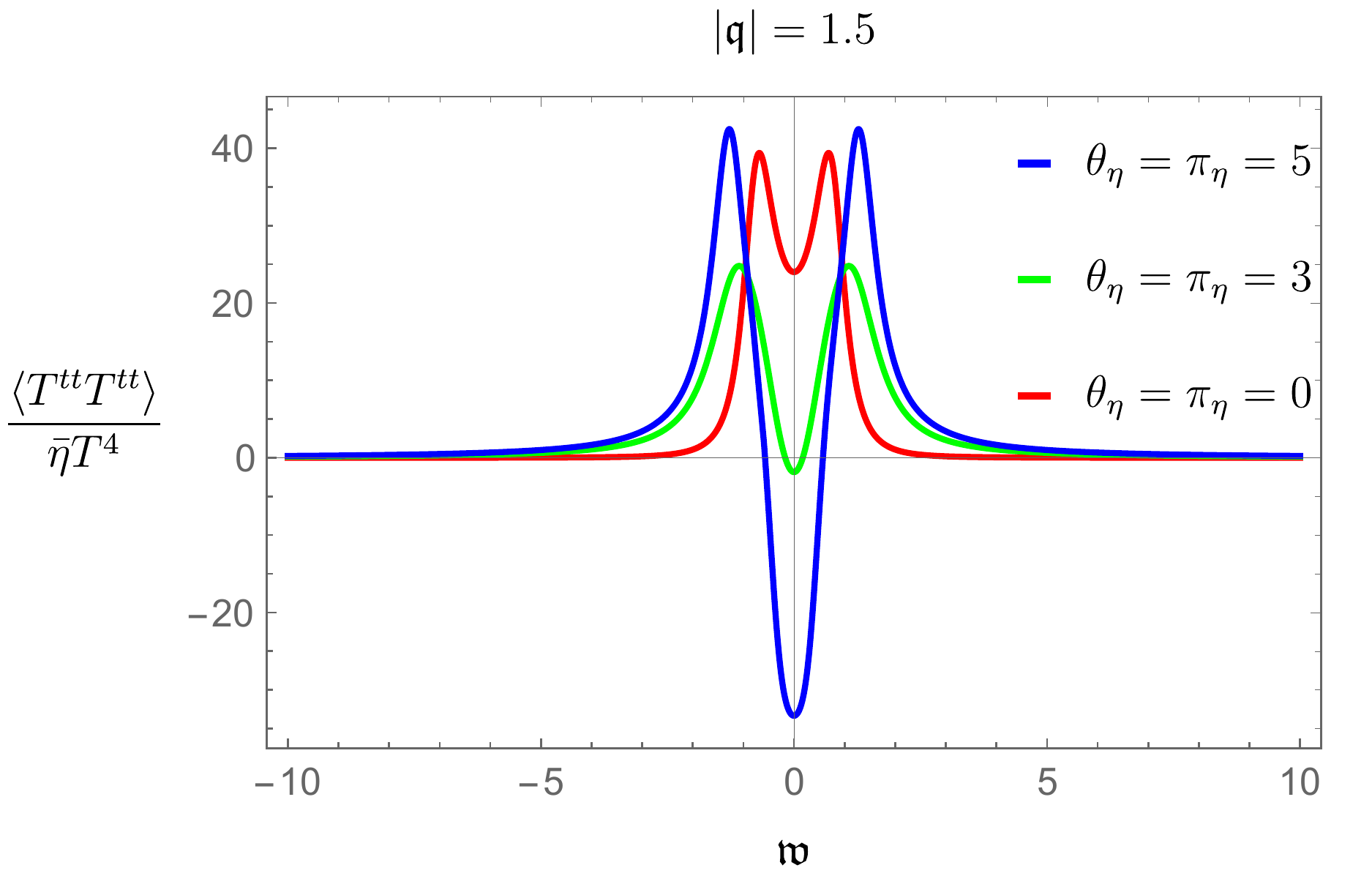}
		\caption{Energy density correlation function in the Landau-Lifshitz frame (red), in an unstable frame (green), and in a stable frame (blue). }
		\label{TT}
	\end{figure}
	Solving $	\mathcal{D}_L(\wn, \qn)=0$ gives the spectrum of modes associated with the longitudinal channel. This has been discussed in detail in \cite{Kovtun:2019hdm}. However, we illustrate the pole structure of correlation functions associated with the sound channel in figure \eqref{pole_sound}.
	As it is seen, the correlation function  develops a range of negative values in unstable frames. The situation shown in the figure is for $\qn=1.5$. The smaller the value of $\qn^2$, the narrower the range of negative values around $\wn=0$. Close to $\wn=0$ we have
	\begin{equation}\label{limit}
		\lim_{\wn\rightarrow0}G_{T_{tt}T_{tt}}(\wn, \qn)=-\frac{6\,\bar{\eta}\,T^4\,\big(\pi_{\eta}^2\theta_{\eta}\,\qn^2+48(\pi_{\eta}-4)\big)}{\theta_{\eta}(\pi_{\eta}-4)\,\qn^2+48}\,.
	\end{equation}
	As it is clearly seen, in a stable frame ($\theta_{\eta}\,,\pi_{\eta}>4$) \footnote{The sufficient conditions for stability and causality, discussed earlier, take a simpler form when expressed in terms  of dimensionless coefficients: $\pi_{\eta}>4$ and $\theta_{\eta}>4$.}, \eqref{limit} is negative for any value of $\qn^2$ \footnote{However, for unstable frames ($\theta_{\eta}=\pi_{\eta}<4$), this negative range may disappear depending on how small $\qn$ is.}.  Of course, although $G_{T_{tt}T_{tt}}$ has such a negative value interval, it does not mean that the correlation function of a hermitian operator can become negative, but that the relativistic hydrodynamics in the general frame has special features that lead to this result. Below, we explain what exactly these features are.
		\vspace{2mm}
		
We can track this negativity by looking at \eqref{G_TT_TT}. At small $\wn$ the first line is dominant; in  a stable frame, the largest negative contribution to $G_{T_{tt}T_{tt}}$ at small $\qn$ then comes from the last term: $\sim (-6 \bar{\eta}T^4 \qn^2)\times2304(\pi_{\eta}-4)$. The latter itself, is coming from $\big\langle\frac{\delta T}{T}\frac{\delta T}{T}\big\rangle_{\wn\qn}$ in \eqref{G_first}. This temperature correlator is calculated in appendix \ref{hydro_fluc_App}.  There it is clearly seen that the expression $(\pi_{\eta}-4)$ arises because the coefficient of $\qn^2$ in the second line of \eqref{Long_Channel_eq_set_1} has the same factor. This is precisely because the coefficient of $\boldsymbol{\nabla}^2\delta \textbf{v}$ in the momentum conservation equation \eqref{EoM_z} also has this factor. 
	\vspace{2mm}
		
We see that the specific form of the BDNK equations used to solve the causality and stability problems leads to the behavior observed in the figure \ref{TT}. More precisely, not imposing the first derivative on-shell condition, i.e. the vanishing of the underlined expressions inside the brackets in \eqref{EoM_t} and \eqref{EoM_z}, is necessary to ensure the stability of the equilibrium; however, according to the discussion in the previous paragraph, there is also another result; it will cause the energy density correlation function to be negative at the \textit{small frequency} limit \footnote{Note that if we impose on-shell conditions on the equations \eqref{EoM_t} and \eqref{EoM_z}, the two coefficients $\bar{\theta}$ and $\bar{\eta}$ will be washed out of the equations. Therefore, we end up with the familiar equations in the Landau-Lifshitz frame. This is equivalent to taking $\theta_{\eta}=\pi_{\eta}=0$ in \eqref{limit}; obviously, $ \lim_{\wn\rightarrow0}G_{T_{tt}T_{tt}}(\wn, \qn)= 24 \bar{\eta}T^4$, which is a positive number.}.Interestingly, this feature does not appear in the\textit{ small momentum} limit when the frequency is held fixed.
		\vspace{2mm}
		
%

Now, let us study the analytic structure of the correlation functions. This can be done by finding the roots of $\mathcal{D}_{L}(\wn,\wn)=0$ (see \eqref{D_L}). The corresponding pole structure is shown in Figure \ref{pole_sound}. As shown on the right panel of the figure, there are eight pole singularities in the complex $\wn$ plane at any particular value of $\qn$ (indicated by a specific color). Four of them located in the lower half plane correspond to retarded correlators, while four located in the upper half plane correspond to advanced correlators. The momentum range in which the BDNK spectrum is consistent with conventional relativistic hydrodynamics (shown in the left figure) is identified by the dashed circle in the right panel. We find that for $\theta_{\eta}=4$ this is given by $\qn\lesssim0.5$. We see that from the four modes of the BDNK spectrum in the lower half-plane two of the modes lie outside this range, in agreement with \cite{Kovtun:2019hdm}.
	\subsection{Correlation functions in  transverse channel}
	\label{transverse}
\begin{figure}[tb]
	\centering
	\centering
	\includegraphics[width=0.48\textwidth]{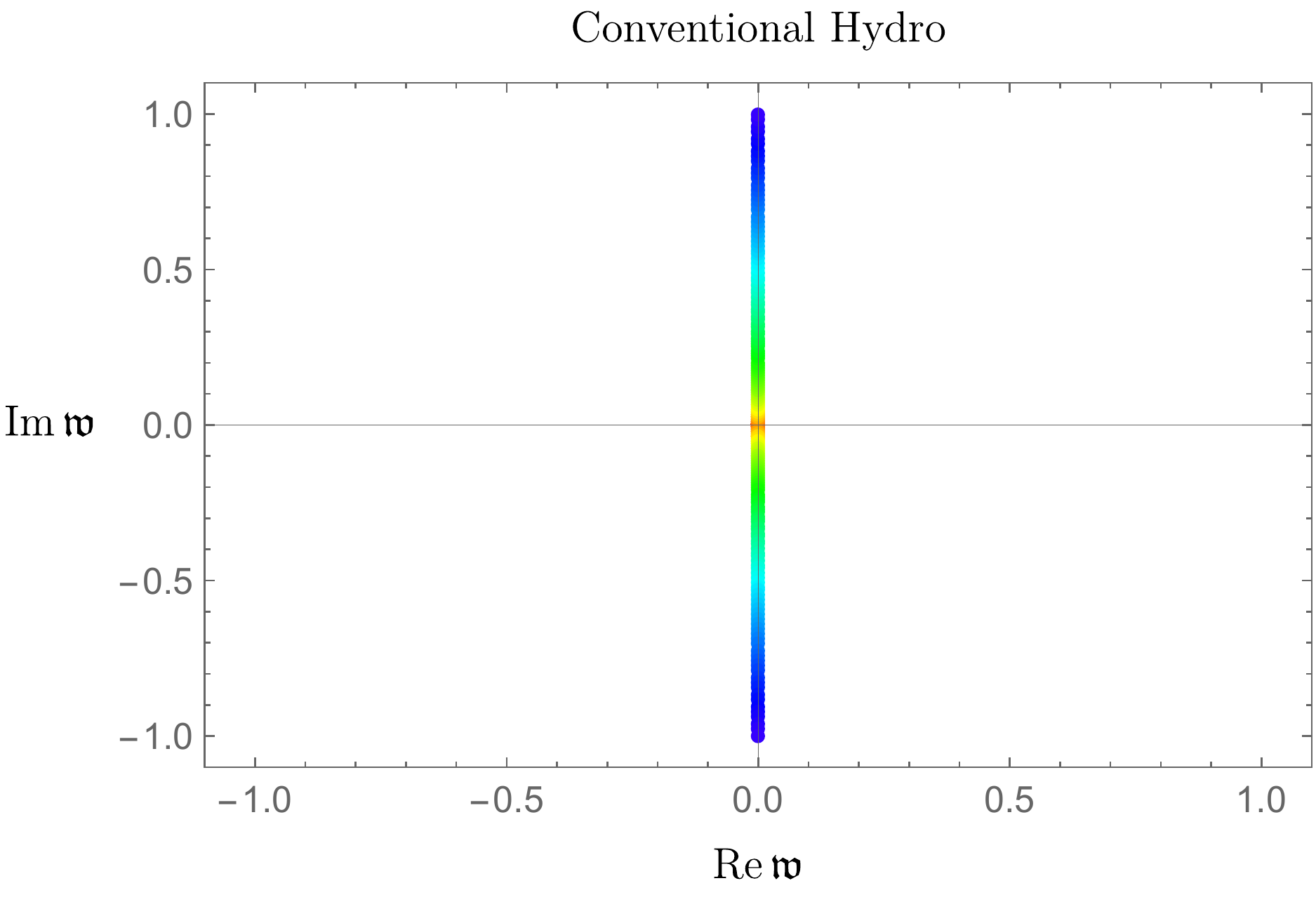}\,\,\,\,\,\includegraphics[width=0.48\textwidth]{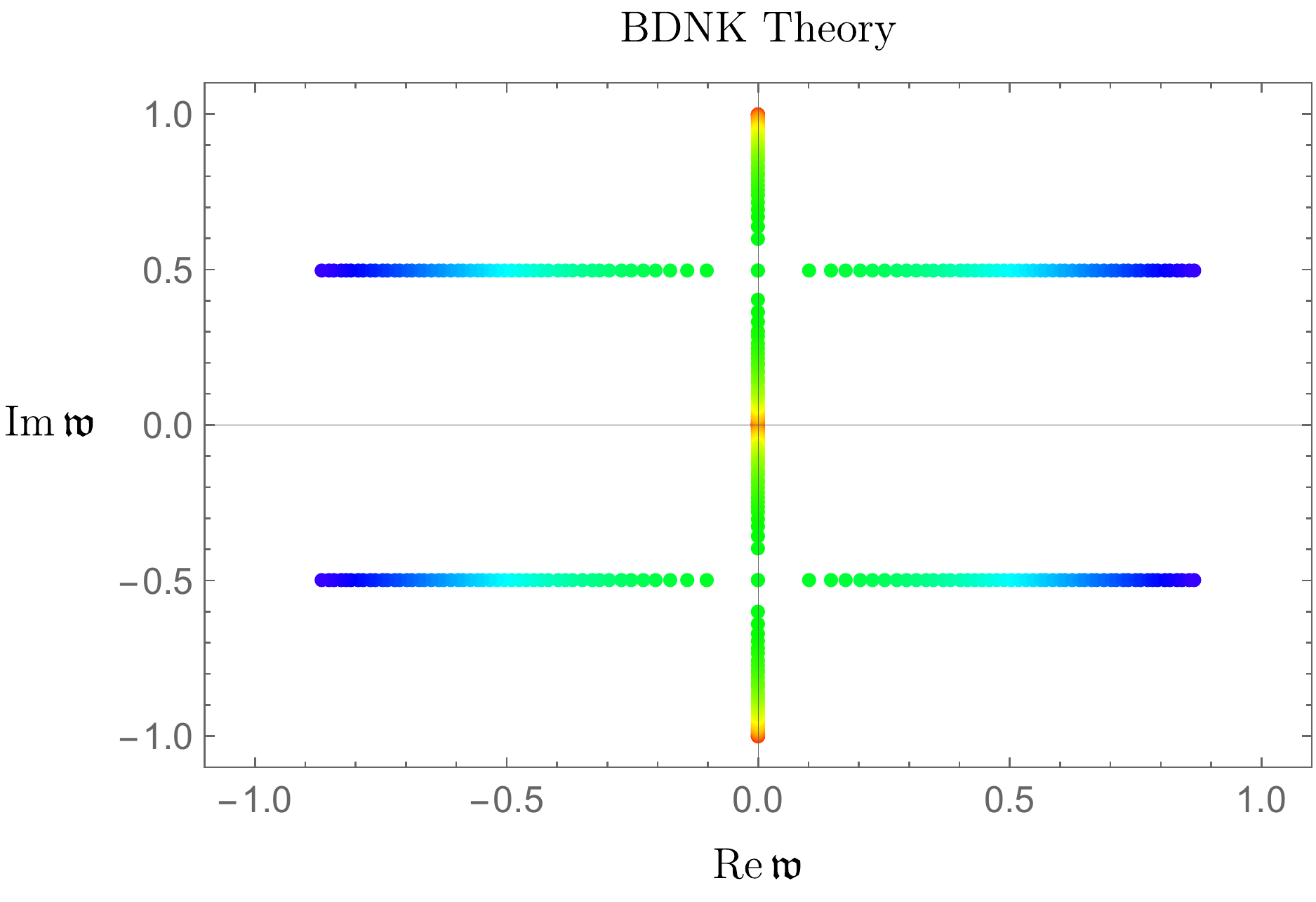}\,
	\caption{Correlation functions poles comparison between conventional relativistic hydrodynamics and BDNK theory in shear channel. Each colorful trajectory starting in red and ending in purple illustrates change in one single mode when $\qn$ varies from $0$ to $2$. Lower half plane modes corresponds to poles of retarded Green's functions in shear channel, while the upper half plane modes corresponds to the poles of advanced Green's functions. Conventional hydro Green's functions have only one single shear modes in lower half plane while BDNK theory Green's functions have one more extra modes associated with the inclusion of the transport coefficient $\theta$.  }
	\label{pole_shear}
\end{figure}

	Among various correlation functions associated with this channel, we choose to explicitly express $G_{T_{ti_{\perp}}T_{ti_{\perp}}}(\wn,\qn)=\langle T^{ti_{\perp}}T^{ti_{\perp}} \rangle_{\wn \qn}$, $i=x,y$. To this end, we first use the energy momentum tensor at linear order and write
	\begin{equation}\label{}
		G_{T_{ti_{\perp}}T_{tj_{\perp}}}(\wn,\qn)=\delta_{ij}\, \bar{p}^2T^8 \bigg( \theta_{\eta }^2\wn^2+16\bigg)\langle\delta v_{i_{\perp}}\delta v_{j_{\perp}}\big\rangle_{\wn\qn}
	\end{equation}
	Note that we have set the equilibrium fluid velocity to be $u^{\mu}=(1, 0)$. 
	Then by utilizing the correlation function of the transverse components of the velocity, found in Appendix \ref{hydro_fluc_App}, we arrive at
	\begin{equation}\label{G_Tx_Tx}
			G_{T_{ti_{\perp}},T_{tj_{\perp}}}=\,\delta_{ij}\frac{2\bar{\eta}T^4\,\qn^2 }{|\mathcal{D}_T(\wn, \qn)|^2}  \bigg( \theta_{\eta }^2\wn^2+16\bigg)\\
	\end{equation}
	where $\mathcal{D}_T$ is the spectral function of the transverse (shear) channel  :
	\begin{equation}\label{on_shell}
			\mathcal{D}_T(\wn,\qn)=\theta_{\eta}\wn^2+4 i\wn-\qn^2
	\end{equation}
	Solving $	\mathcal{D}_T(\wn, \qn)=0$ gives the spectrum of modes associated with the transverse  (shear) channel. This has been discussed in detail in \cite{Kovtun:2019hdm}. However, we illustrate the pole structure of correlation functions associated with this channel in figure \eqref{pole_shear}.
In conventional hydrodynamics, there is only one shear mode above and below the complex $\wn$ plane, whereas in BDNK theory there is an additional mode for each half-plane \footnote{This is agreement with \cite{Kovtun:2019hdm}.}. The latter is due to the inclusion of the UV regulator $\theta_{\eta}$ in the theory.
	\section{Correlation functions beyond the linear regime}
	\label{non_linear}
	Hydrodynamics is essentially a nonlinear theory. Nonlinearities manifest as the interactions between hydrodynamic modes. A systematic approach for accounting these interactions is to construct a hydrodynamic effective field theory \cite{Crossley:2015evo,Jensen:2017kzi,Haehl:2015uoc,Liu:2018kfw}. Then from this EFT, the corrected correlation function can be found \cite{Chen-Lin:2018kfl}, showing a large renormalization of the transport coefficients \cite{Kovtun:2011np} as well as the long time tails \cite{Kovtun:2003vj}.
		\vspace{2mm}
		
	These effects can also be explained by some general hydrodynamic statements \cite{An:2019osr}. Let us recall that hydrodynamic variables are macroscopically averaged values of densities over regions of size $b$, where $b\gg \ell_{mic}$. Typically, the microscopic length scale $\ell_{mic}$ is of order of the correlation length: $\ell_{mic}\sim\xi$. In other words, a hydrodynamic fluctuation, such as $\delta v$, can be regarded as averaging over $\big(b/\xi\big)^3$ independent correlation volumes. According to the central limit theorem, the distribution of such average will approach a Gaussian as $\big(b/\xi\big)^3\gg1$. The distribution is Gaussian, the non-Gaussian contribution is suppressed by an additional factor of $\big(b/\xi\big)^{-3}$, and knowledge of the two-point function at the level of the linear response is sufficient to find correlation functions beyond the linear response \cite{Kovtun:2003vj}. 
			\vspace{2mm}
			
Following the above logic, we would like to investigate in this section the analytic structure of a certain correlation function beyond the linear response regime. This includes specifying the threshold singularity and branch cut structure of the associated response function. We emphasize that we do not construct an effective field theory for this, but simply use the assumption that the fluctuation distribution is Gaussian.
	\subsection{Correlation function of the shear stress}
	To reduce technical complexity, we restrict the analysis to the correlation function of the shear stress, $G_{T_{xy}T_{xy}}=\big\langle T_{xy}T_{xy}\big \rangle$. The reason for this selection is as follows.  At linear order, the shear stress $T_{xy}$ vanishes; it just starts to contribute in quadratic order:
	\begin{equation}
		T_{xy}=\,4\bar{p}T^4\,\delta v_x \delta v_y\,;
	\end{equation}
	so the correlation function $G_{T_{xy}T_{xy}}=\big\langle T_{xy}T_{xy}\big \rangle$ is given by (n.l. stands for nonlinear)
	\begin{equation}
		G^{(\text{n.l.})}_{T_{xy}T_{xy}}(t,\textbf{x})=\,w^2\,\big\langle \delta v_x(t,\textbf{x}) \delta v_y(t,\textbf{x})\delta v_x(0,\textbf{0}) \delta v_y(0,\textbf{0}) \big\rangle\,,
	\end{equation}
	with $w=4\bar{p}T^4$. Since the distribution of fluctuations is assumed to be Gaussian, the above correlation can be factorized as 
	\begin{equation}
		G^{(\text{n.l.})}_{T_{xy}T_{xy}}(\omega,\textbf{k})=w^2\int\frac{d\omega'}{2\pi}\int\frac{d^3k'}{(2\pi)^3}\,G_{\delta v_x\delta v_x}(\omega',\textbf{k}')G_{\delta v_y\delta v_y}(\omega-\omega',\textbf{k}-\textbf{k}')\,.
	\end{equation}
	The two correlation functions in the integrand are  those already calculated in the linear response regime. It is now clear why we chose to use shear stress: this is actually a special case, and to study it the only information about the linear response we need is the correlation function of velocity in the `transverse channel' (see Appendix \ref{App_trans}).
		\vspace{2mm}
		
	Taking $p=(\omega,\textbf{k})$, the integrand can be written as
	\begin{equation}
		\bigg(\underbrace{G^{(+)}_{\delta v_x\delta v_x}(p')}_{\textcircled{1}}+\underbrace{G^{(+)}_{\delta v_x\delta v_x}(-p')}_{\textcircled{2}}\bigg)\bigg(\underbrace{G^{(+)}_{\delta v_y\delta v_y}(p-p')}_{\textcircled{3}}+\underbrace{G^{(+)}_{\delta v_y\delta v_y}(-p+p')}_{\textcircled{4}}\bigg)
	\end{equation}
	where $(+)$ represents the one-sided Fourier transformation (see Appendix \ref{hydro_fluc_App} for details). Of the four possible multiplicative terms, $\textcircled{1}\times\textcircled{4}$ and $\textcircled{2}\times\textcircled{3}$ do not contribute to the integral. The reason is simply that, for each of them, all pole singularities lie in only half of the complex $\omega'$ plane; either in the upper half or in the lower half. Regarding the remaining two terms, $\textcircled{2}\times\textcircled{4}$ only contributes to the upper half plane structure of $G^{(\text{n.l.})}_{T_{xy}T_{xy}}(p)$, the advanced Green's function. And finally this is $\textcircled{1}\times\textcircled{3}$ term which contributes to the lower half plane structure of $G^{(\text{n.l.})}_{T_{xy}T_{xy}}(p)$. Therefore, to find the analytic structure of the ``response function'' of $T_{xy}$, i.e., $G^{R}_{T_{xy}T_{xy}}$, it is sufficient to consider the following integral
	\begin{equation}\label{int_plus_plus}
		\begin{split}
			\bar{w}^2&\int\frac{d\omega'}{2\pi}\int\frac{d^3k'}{(2\pi)^3}\,G^{(+)}_{\delta v_x\delta v_x}(\omega',\textbf{k}')G^{(+)}_{\delta v_y\delta v_y}(\omega-\omega',\textbf{k}-\textbf{k}')\\
			&\sim \int\frac{d\wn'}{2\pi}\int\frac{d^3q'}{(2\pi)^3}\,\frac{\mathcal{N(\wn')}}{\mathcal{D}_{T}(\wn',\qn')}\frac{\mathcal{N(\wn-\wn')}}{\mathcal{D}_{T}(\wn-\wn',\qn-\qn')}\,.
		\end{split}
	\end{equation}
	The expression $\mathcal{D}_T$ in the denominator is the spectral function of the transverse channel that we already found  in the linear response regime (see \eqref{on_shell})
	\begin{equation*}\label{}
			\mathcal{D}_{T}\big(\wn,\qn)=\textbf{0}\big)=\theta_{\eta}\wn^2+4 i \wn-\qn^2=0
	\end{equation*}
	and encodes the dispersion relations of the two gapped modes associated with this channel
	\begin{equation}\label{on_shell_D_T}
		\wn_{1,2}=\,-\frac{i}{\theta_{\eta}}\big(2\mp\sqrt{4-\theta_{\eta}\qn^2}\big)\,.
	\end{equation}
	Note that we are considering fluctuations on top of an equilibrium state with $u^{\mu}=(1,\textbf{0})$.
		\vspace{2mm}
		
	Needless to say, the integral \eqref{int_plus_plus} is a kind of one-loop integral that we usually deal with in quantum field theory:
	\begin{figure}[!ht]
		\centering
		\begin{tikzpicture}
			[decoration={markings,
				mark=at position 3cm with {\arrow[line width=1pt]{>}},
			}
			]
			\draw[help lines,line width=0.8pt] (-2,0) -- (-.6,0) coordinate (xaxis);
			\draw[help lines,line width=0.8pt] (.6,0) -- (2,0) coordinate (xaxis);
			\draw (0,0) node[minimum size=1.2cm,draw,circle] {};
			\node[] at (-1.5,.3) {$ \big(\wn, \qn\big)$};
			\node[] at (0,1) {$ \big(\wn', \qn'\big)$};
		\end{tikzpicture}
	\end{figure}
	\\It is convenient to use Feynman parameters to rewrite  \eqref{int_plus_plus} as 
	\begin{equation}\label{Feynman_parameter}
		\mathcal{I}(\wn,\qn)=\int\frac{d\wn'}{2\pi}\int\frac{d^3q'}{(2\pi)^3}\,\int_{0}^{1}d\alpha_1\int_{0}^{1}d\alpha_2\,\delta(\alpha_1+\alpha_2-1)\,\frac{\mathcal{N(\wn')}N(\wn-\wn')}{\big[\alpha_1\mathcal{D}_{T}(\wn',\qn')+\alpha_2 \mathcal{D}_{T}(\wn-\wn',\qn-\qn')\big]^2}
	\end{equation}
	Then based on general complex analysis arguments, it is possible to discuss the analytic properties of $\mathcal{I}(\wn,\qn)$, without performing the frequency/momentum integral. This is basically the idea proposed by Landau in the famous paper \cite{Landau:1959}. Here, our goal is to use the Landau conditions/equations derived in \cite{Landau:1959} to find poles and branch cuts of $\mathcal{I}(\wn,\qn)$ in \eqref{Feynman_parameter}, which are actually the singularities of $G^{R}_{T_{xy}T_{xy}}$ as well.
	\subsection{Analytic structure of response functions from ``Landau conditions/equations''}
	\label{reason}
	 A physical question is why we should be interested in specifying the analytic structure of the response functions,  while we do not have access to their explicit expressions. This can be discussed in the context of quantum field theory. In QFT, the discontinuity of the general scattering amplitude $\mathcal{M}$ can be found by using the Cutkosky algorithm	\cite{Schwartz:2014sze}\footnote{Note that $\mathcal{I}(\wn,\qn)$ above is similar to $\mathcal{M}$; although there is no any asymptotic state in the shear channel to define scattering amplitude \cite{Endlich:2010hf}.}. This is done by letting all internal lines in a Feynman diagram go on-shell. Knowledge of threshold singularities and branch cuts is required at this point to perform the final Feynman integration \footnote{A threshold singularity is the branch point of the scattering amplitude $\mathcal{M}$. See Appendix \ref{Landau_App} for more details.}. From this, $\text{Im} \mathcal{M}$ can be calculated. This might then be used to find some specific decay rate or cross section via ``optical theorem". These processes can be performed in any order in perturbation theory \cite{Schwartz:2014sze}.
	 	\vspace{2mm}
	 	
          Our motivation for studying the analytic structure of the hydrodynamic response function is actually based on the discussion in the previous paragraph. In the context of hydrodynamics, QFT processes can be important in a variety of situations. In particular, scattering of phonon and vortex excitations  is extensively discussed in the literature 			\cite{Endlich:2010hf,Gripaios:2014yha}. What we will do in the next two subsections is to bring some QFT methods into the context of stable first order relativistic hydrodynamics for the first time and explore the analytic structures. These methods are not only useful in the case of BDNK theory, but also simplify loop calculations in EFT of hydrodynamics.
          	\vspace{2mm}
                
We first develop field-theoretic methods to specify the branch cut structure of the shear stress response function, $G^{R}_{T_{xy}T_{xy}}$, in \sec{branch_cut}.
The threshold singularity of $G^{R}_{T_{xy}T_{xy}}$ can be found by applying the Landau condition to $\mathcal{I}(\wn,\qn)$ (see appendix \ref {Landau_App} for a brief review). This is what will be done in \sec{threshold}. We leave the use of the results in the QFT processes to future work.
	\subsubsection{Branch cuts from on-shell conditions}
	\label{branch_cut}
In the case of Feynman  integral \eqref{int_plus_plus} all of the kinematic dependence is encoded in the integrand. The branch cuts of such integral is uniquely specified by where the integrand is singular on the integration contour\cite{Hannesdottir:2022xki}. The latter is equivalent to requiring
	\begin{equation}\label{conditions}
		\mathcal{D}_{T}(\wn',q')=0\,,\,\,\,\,\,\,\, \mathcal{D}_{T}(\wn-\wn',q\pm q')=0\,.
	\end{equation}
	These are actually on-shell conditions coming from the denominator of \eqref{int_plus_plus}. Note that we have used  $\int\frac{ d^3q'}{(2\pi)^3}=\frac{1}{(2\pi)^2}\int q'^2dq'\int_{-1}^{1}d\cos \theta'$ and performed the integration over $\theta'$, which is the angle between vectors $\qn$ and $\qn'$. Let us recall that the only momentum dependence in \eqref{on_shell} is a $\qn^2$ term. Thus $\mathcal{D}_{T}(\wn-\wn',q\pm q')$ depends on $\theta$ through $q^2+q'^2-2 q q'\cos\theta $. Among other things,  integration over $\theta$ gives a factor of 
	\begin{equation}\label{}
	\ln \bigg( \mathcal{D}_T\big(\wn-\wn',q^2+q'^2-2 q q'\cos\theta \big)\bigg)\bigg|^{\cos \theta=1}_{\cos \theta=-1}.
		\end{equation}
Singularity caused by this factor is what has been given in the second equation in \eqref{conditions}.
		\vspace{2mm}
		
	Our goal in this section is to solve the conditions \eqref{conditions}. Let us assume that the entire solution is identified by the following equation
	\begin{equation}\label{singularity}
		s(\wn, \qn)=0\,.
	\end{equation}
	For a given $\qn$, this equation represents  a set of points in the complex $\wn$ plane: the set $\mathcal{S}_{\qn}$. 
	\textit{In the following, we introduce a method to solve \eqref{conditions}, not for the function $s$, but directly for the set $\mathcal{S}_{\qn}$.}
		\vspace{2mm}
		
	Clearly, the two equations in \eqref{conditions} are the on-shell conditions for the two lines of the loop when the internal momenta are parallel ($\cos\theta =\pm1$).
	This simply means that for a given external $(\wn,\qn)$, each line of the loop carries one of the two modes \eqref{on_shell_D_T}, in $\qn'||\qn$ conditions. We then conclude that the set $\mathcal{S}_{\qn}$ must be given by
	\begin{equation}\label{S_qn}
		\mathcal{S}_{\qn}=\bigcup_{\qn'||\qn}\,\,\bigcup_{i,j\in\{1,2\}} \bigg\{\wn_i(\qn')+\wn_j(\qn-\qn')\bigg\}\,.
	\end{equation}
	To make things simpler, we take the external momentum directed in one specific direction: $\qn=(0,0,q)$. Then by reparameterizing the momenta in the loop as
	\begin{equation*}\label{}
		\begin{tikzpicture}
			[decoration={markings,
				mark=at position 3cm with {\arrow[line width=1pt]{>}},
			}
			]
			\draw[help lines,line width=0.8pt] (-2,0) -- (-.6,0) coordinate (xaxis);
\draw[help lines,line width=0.8pt] (.6,0) -- (2,0) coordinate (xaxis);
\draw (0,0) node[minimum size=1.2cm,draw,circle] {};
			\node[] at (-1.5,.3) {$ \big(\wn, \qn\big)$};
			\node[] at (0,1.) {$ \big(\wn_i, \frac{\qn}{2}+\qn'\big)$};
						\node[] at (0,-1.) {$ \big(\wn_j, \frac{\qn}{2}-\qn'\big)$};
		\end{tikzpicture}
	\end{equation*}
	equation \eqref{S_qn} takes the following form
	\begin{equation}\label{S_qn_aplplied}
		\boxed{
			\mathcal{S}_{\qn}=\bigcup_{q'\,\ge\,0}\,\,\bigcup_{i,j\in\{1,2\}} \bigg\{\wn_i\big(\frac{q}{2}+q'\big)+\wn_j\big(\frac{q}{2}-q'\big)\bigg\}}
	\end{equation}
	Regarding the evaluation of $\mathcal{S}_{\qn}$ above, some comments are in order:
	\begin{itemize}
		\item We perform the evaluation of $\mathcal{S}_{\qn}$ in four separate parts, $\mathcal{S}^{ij}_{\qn}; i,j\in\{1,2\}$, corresponding to the various values the pair $(i,j)$ may take. 
		\item At a given $q$, for each part, e.g., $(i,j)=(1,1)$, we calculate $\wn_i\big(\frac{q}{2}+q'\big)+\wn_j\big(\frac{q}{2}-q'\big)$ in a large number of non-negative values of $q'$. 
		\item It turns out that for a particular $(i,j)$, the points $\wn_i\big(\frac{q}{2}+q'\big)+\wn_j\big(\frac{q}{2}-q'\big)$ form a curve with two cusp points in the complex $\wn$ plane.
		\item Depending on $(i,j)$, the points corresponding to $\wn_i\big(\frac{q}{2}+q'\big)+\wn_j\big(\frac{q}{2}-q'\big)$ are accumulated around those associated with either $q'=0$ or $q'\rightarrow \infty$.
	\end{itemize}
	As we will see below, the last two comments above point out to  the branch cut and branch point singularities of $\mathcal{I}$ (or equivalently $G^{R}_{T_{xy}T_{xy}}$), respectively.
	\begin{figure}[tb]
		\centering
		\centering
		\includegraphics[width=0.48\textwidth]{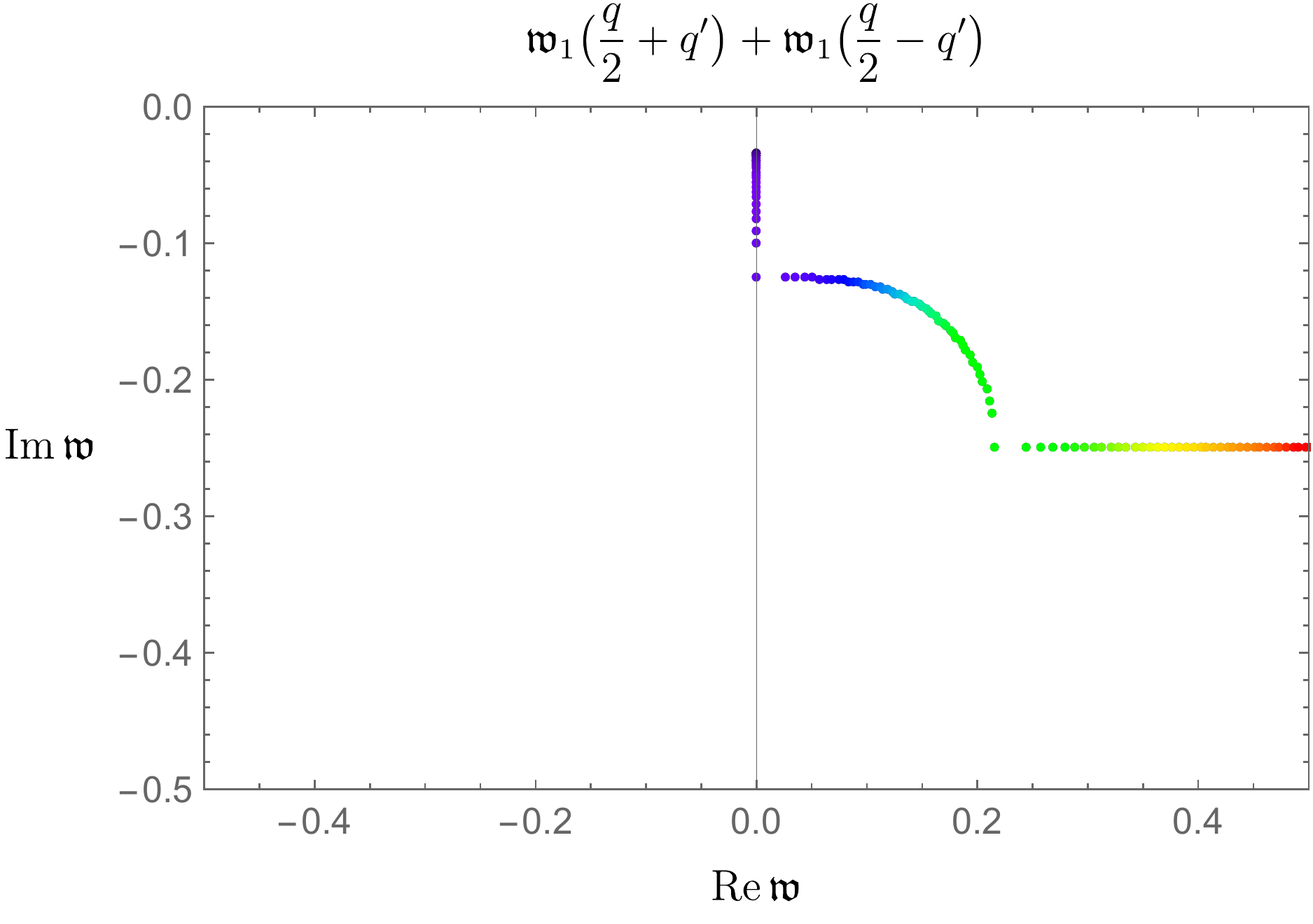}\,\,\,\,\includegraphics[width=0.48\textwidth]{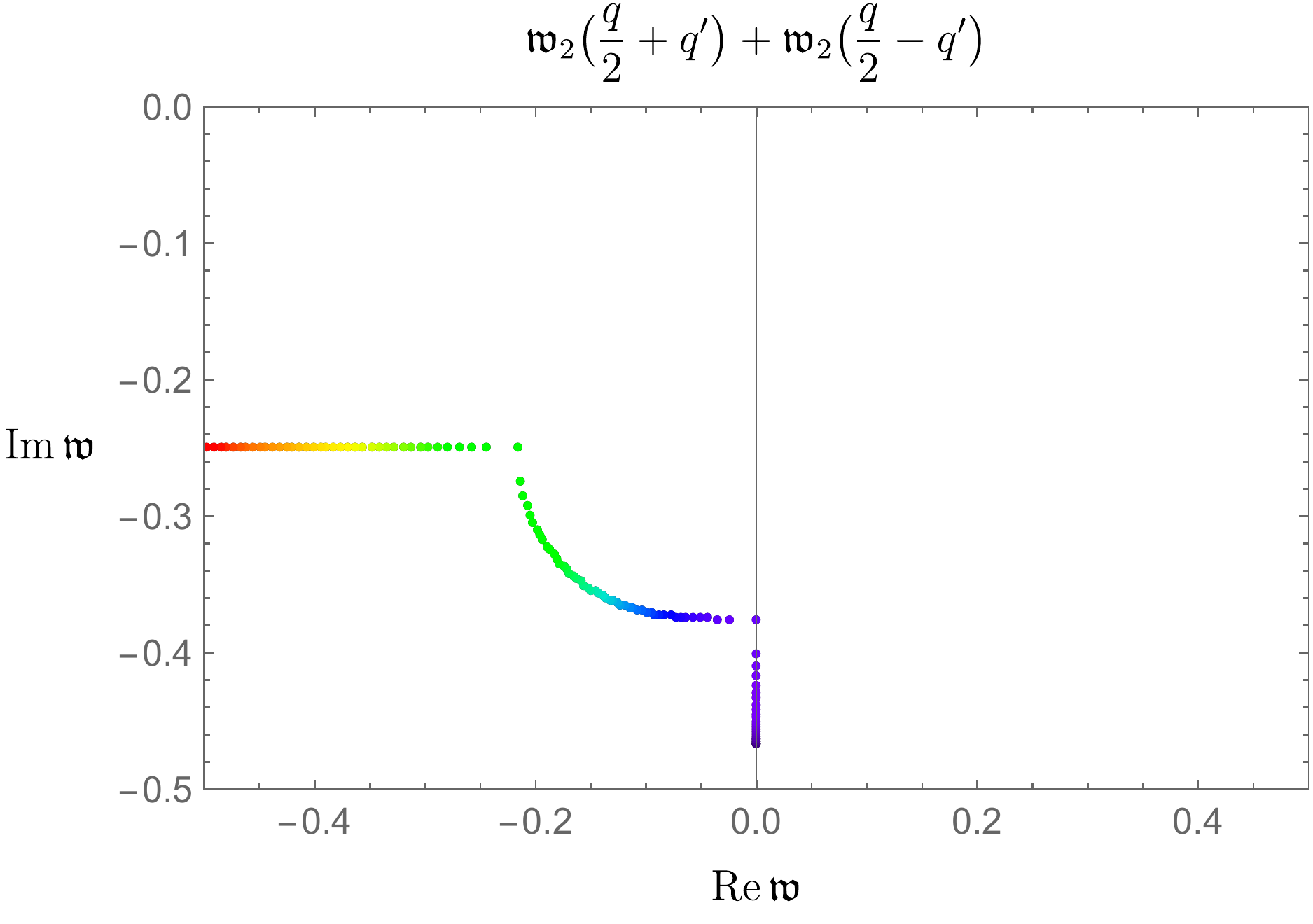}\,
		\includegraphics[width=0.48\textwidth]{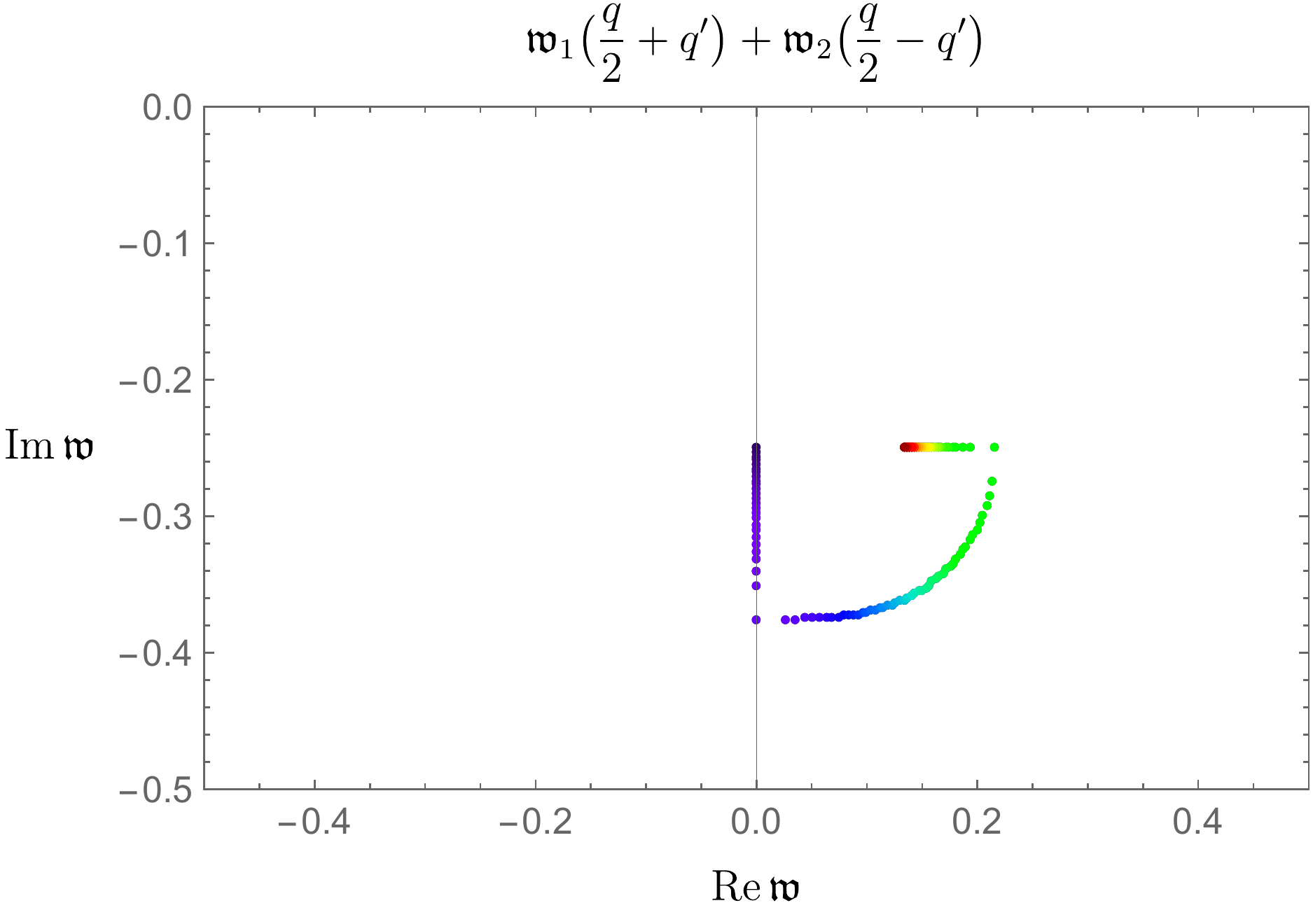}\,\,\,\,\includegraphics[width=0.48\textwidth]{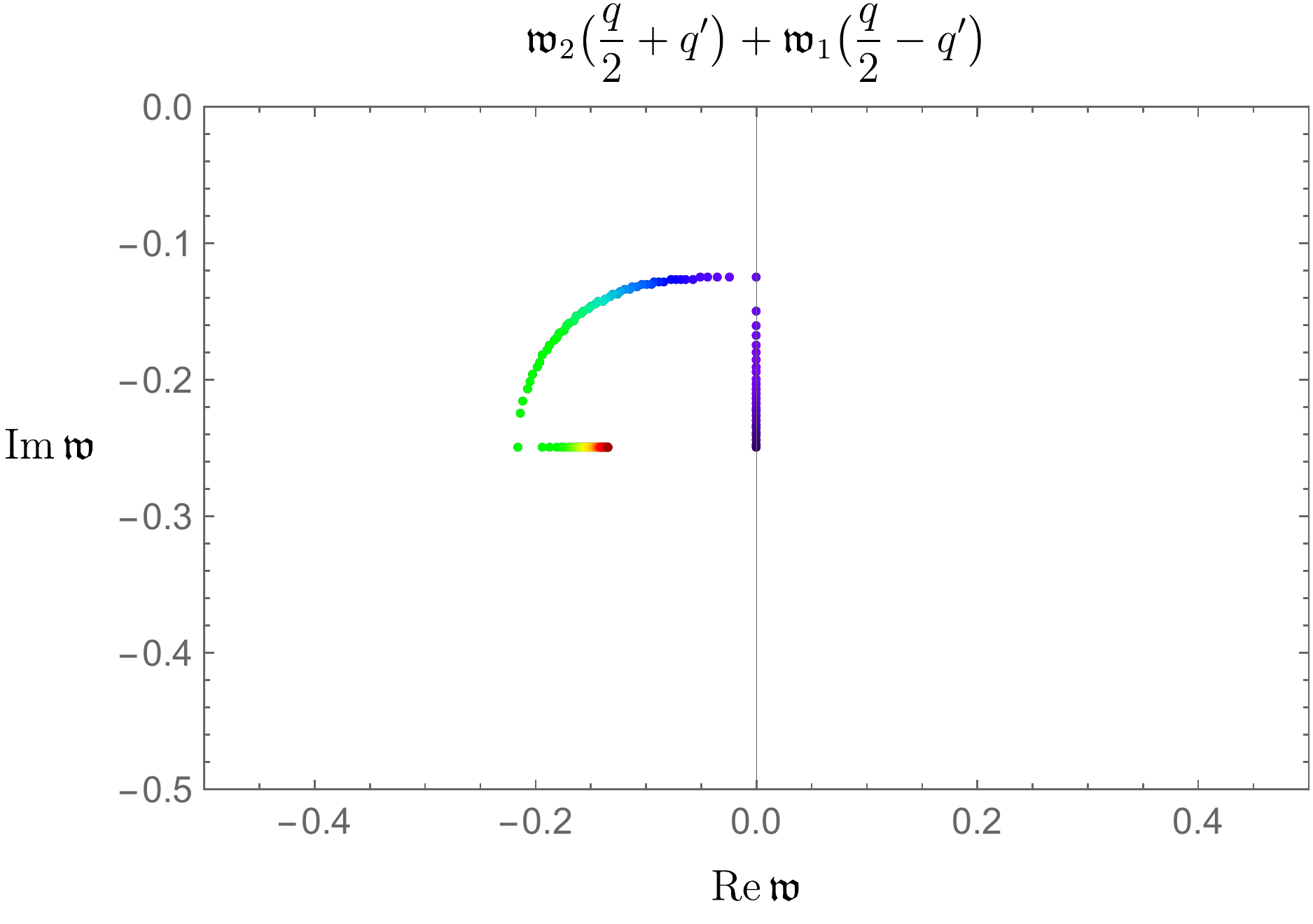}\,
		\caption{Illustration of the $\mathcal{S}^{ij}_{\qn}$'s at $\qn=(0,0,q=0.5)$ in the lower half of the complex frequency plane. In each panel, we have shown values of $\wn_i\big(\frac{q}{2}+q'\big)+\wn_j\big(\frac{q}{2}-q'\big)$ for 150 different $q'$ values between $0\le q'<1.5$. }
		\label{four_parts}
	\end{figure}
	
	\begin{figure}[tb]
		\centering
		\centering
		\includegraphics[width=0.47\textwidth]{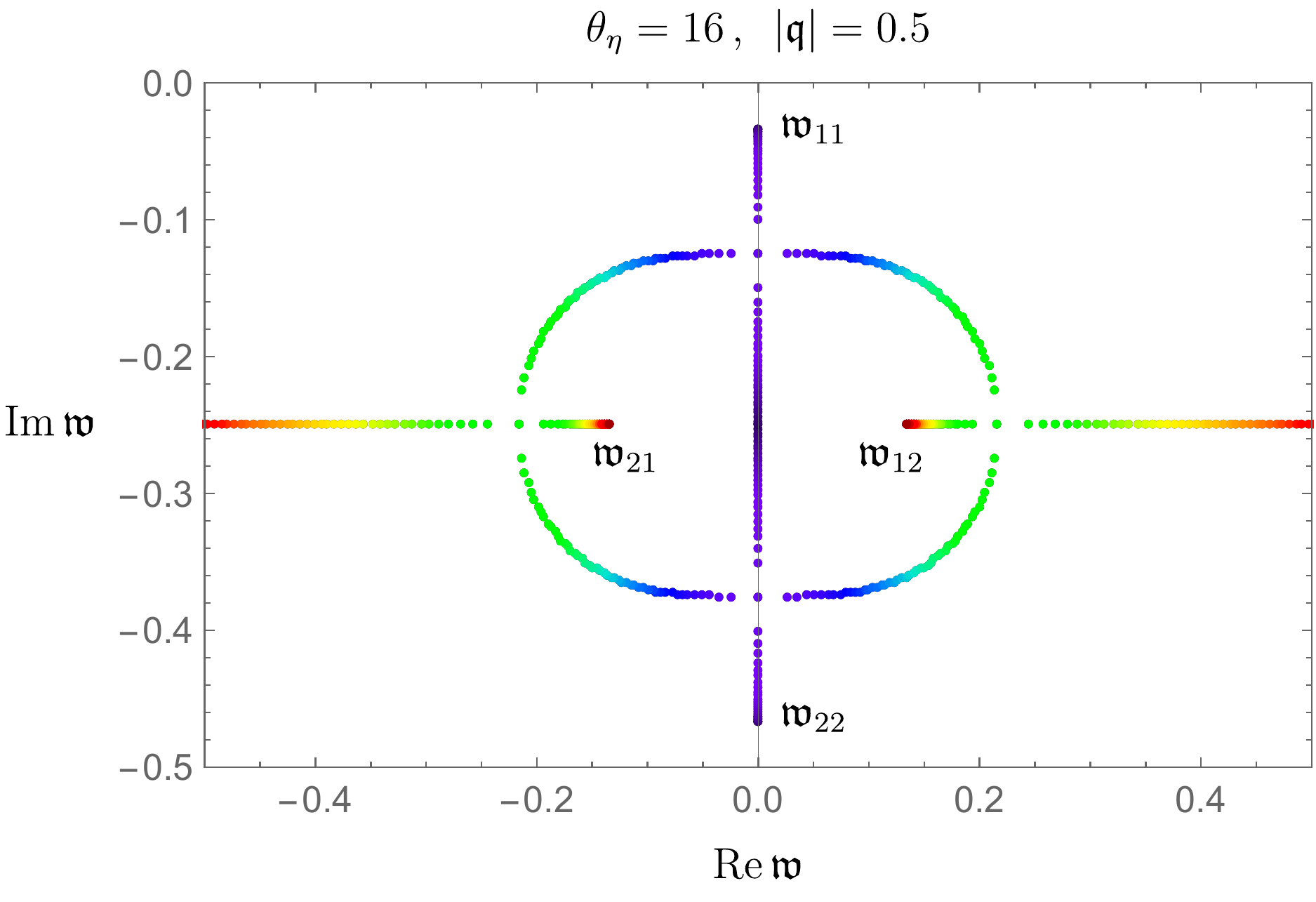}\,\,\includegraphics[width=0.55\textwidth]{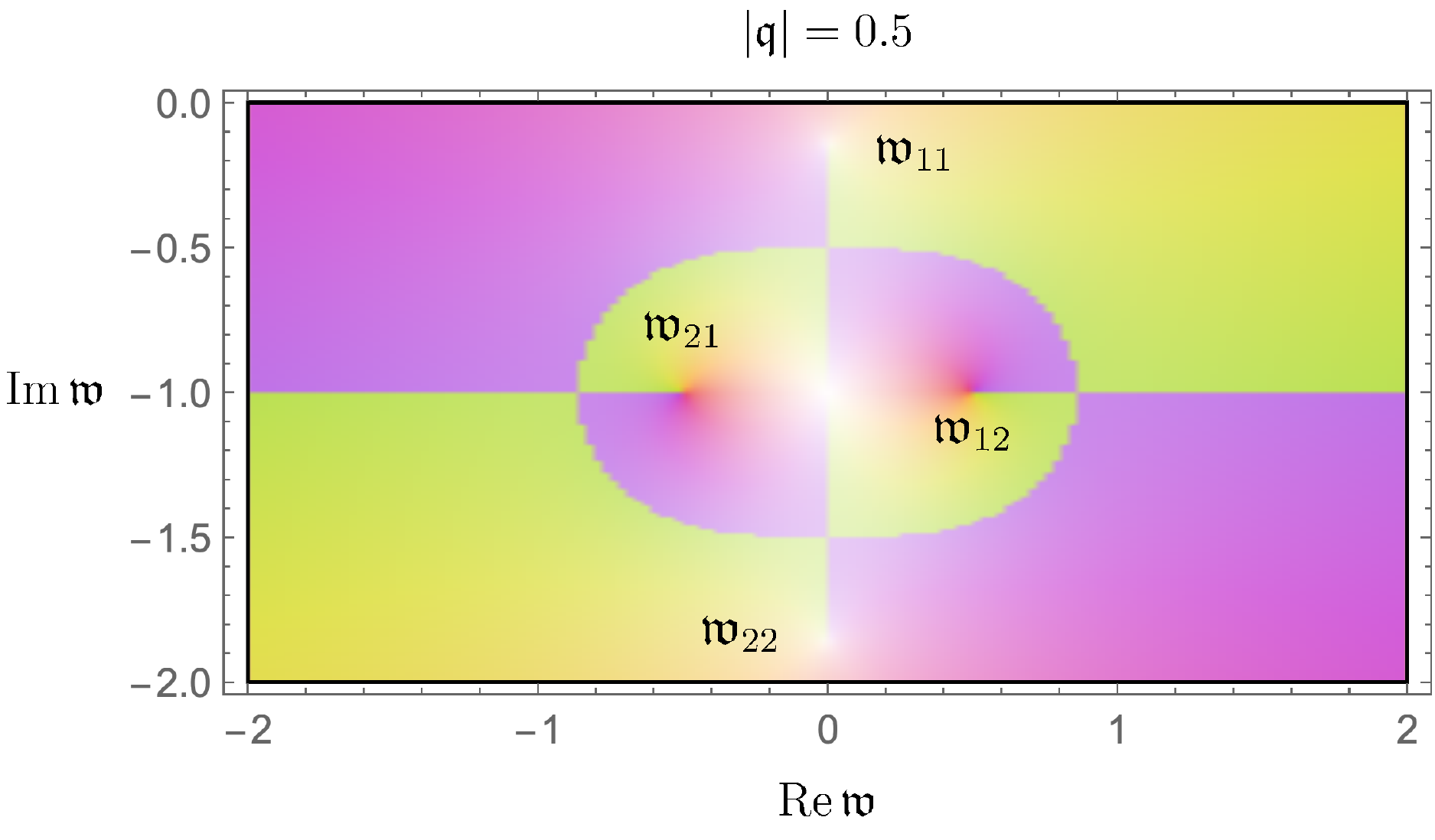}\,
							\includegraphics[width=0.49\textwidth]{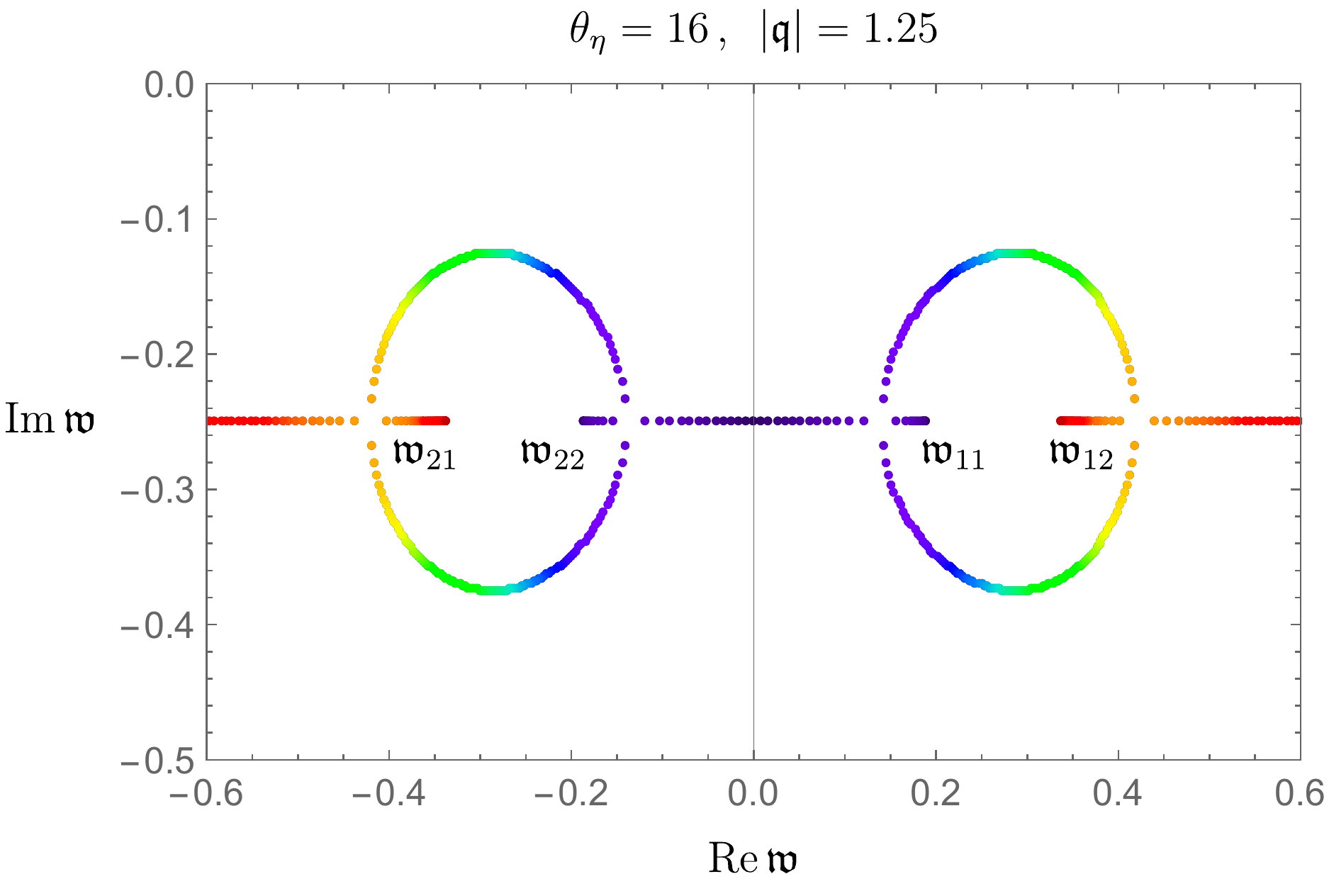}\includegraphics[width=0.55\textwidth]{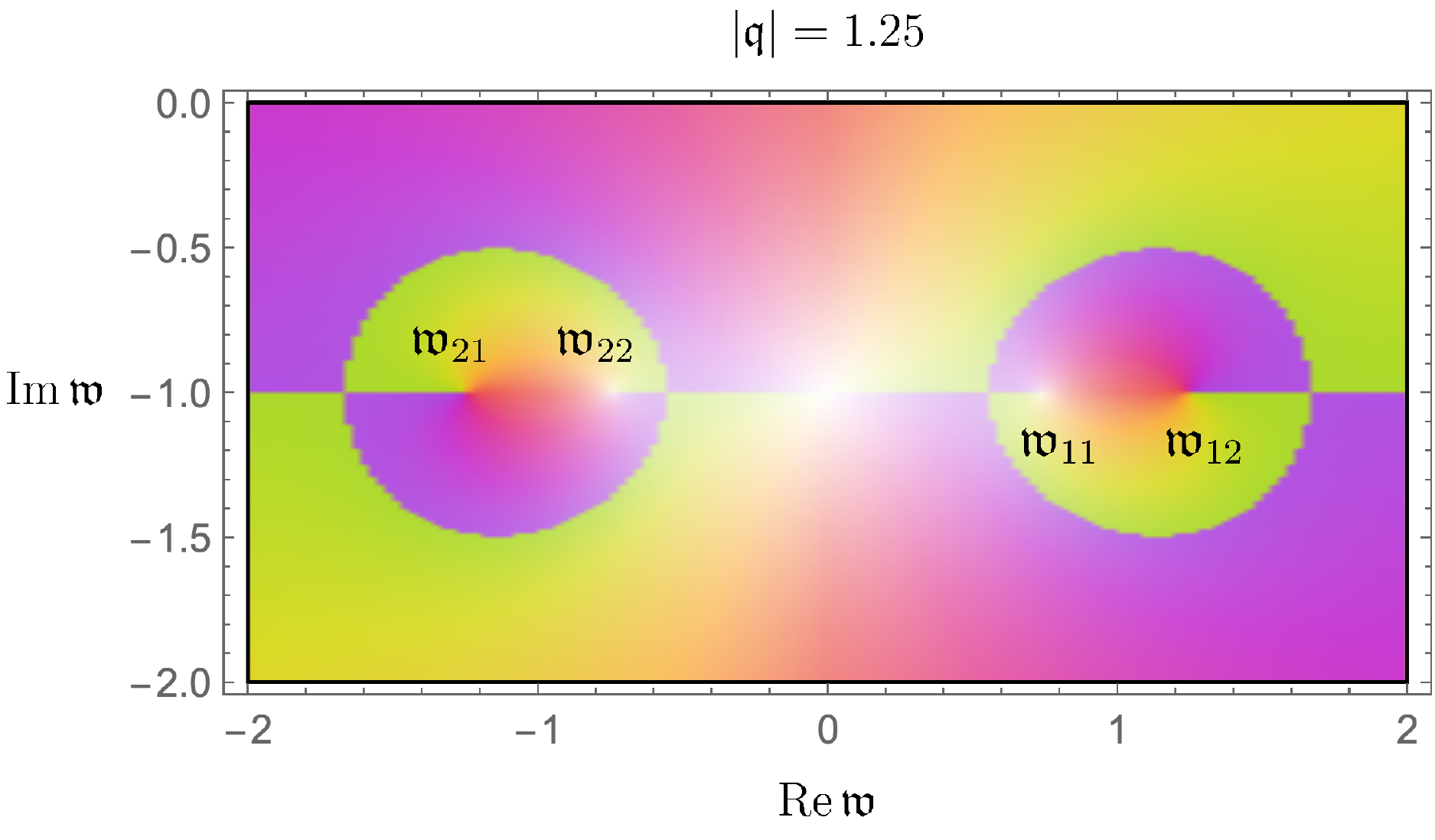}\,
		\caption{Left panel: Branch cut structure of $G^{R}_{T_{xy}T_{xy}}$, found from solving the on-shell equations. Four branch point singularities $\wn_{11}$, $\wn_{22}$, $\wn_{12}$, and $\wn_{12}$ corresponds to the four parts of $\mathcal{S}_{\qn}$, shown in figure  \ref{four_parts}. Right panel: Branch cut structure of $G^{R}_{nn}$, found from explicit loop calculations. The right panel plots are taken from \cite{Abbasi:2022aao}.}
		\label{Compare_before}
	\end{figure}
			\vspace{2mm}
			
	Specializing to $\theta_{\eta}=16$ in \eqref{on_shell_D_T}, we have shown the four parts of $S_{\qn}$, for $\qn=0.5$, in figure \ref{four_parts}. In each panel, the four comments below \eqref{S_qn_aplplied} can be clearly seen.  Each colored trajectory starts in purple at $q'=0$ and ends in dark red at $q'=1.5$. We have checked that by taking large values of $q'$, no significant features are added to these plots.
		\vspace{2mm}
	
	The set of the four parts of $\mathcal{S}_{\qn}$ depicted in figure  \ref{four_parts} have been shown in the top left panel of figure \ref{Compare_before}. This figure actually shows the entire branch cut structure of $G^{R}_{T_{xy}T_{xy}}$, as the complete solution to \eqref{conditions}.  As it is seen, four branch points have been induced: $\wn_{ij}: i,j=1,2$. Each of these points actually corresponds to producing a pair of on-shell excitations $\wn=\wn_{i}(\qn)$ and $\wn=\wn_{j}(\qn)$, in the loop. We show the latter in the next subsection by solving the Landau loop equations analytically.
		\vspace{2mm}
		
	In the same regard, more recently, ref. \cite{Abbasi:2022aao} announced some analytical results on the structure of branch cuts  in  the theory of ``UV-regulated non-linear diffusion''. The on-shell condition in this theory is given by 
	\begin{equation}\label{}
		\tau^2\omega^2+i \tau \omega -\qn^2=0
	\end{equation}
	where $\tau$ is the relaxation time. A comparison between this equation and the on-shell condition in our present case, namely \eqref{on_shell}, shows that the two cases will be the same if one takes
	\begin{equation}\label{vs}
		\boxed{\theta_{\eta}=16\,\,\,\text{in this paper}}\,\,\,\,\,\,\,\text{vs.}\,\,\,\,\,\,\boxed{\tau=4\,\,\,\text{in ref. \cite{Abbasi:2022aao}}}
	\end{equation}
	The (top) left panel of figure \ref{Compare_before} has been already produced for the value of $\theta_{\eta }$ mentioned in \eqref{vs}. In the (top) right panel of figure \ref{Compare_before}, we have shown the branch cut structure of the response function in ref. \cite{Abbasi:2022aao}, at $\tau=4$, and for the same value of $\qn$ taken in the left panel. Note that this reference shows the result in dimensionless frequency $\wn=\tau\omega$. Therefore, for comparing it with the left panel, all numbers on its both axes must be divided by 4. Doing so, \textit{we find that the two panels are in perfect agreement with each other. }
	\vspace{2mm}
	\begin{figure}[tb]
		\centering
		\centering
		\includegraphics[width=0.48\textwidth]{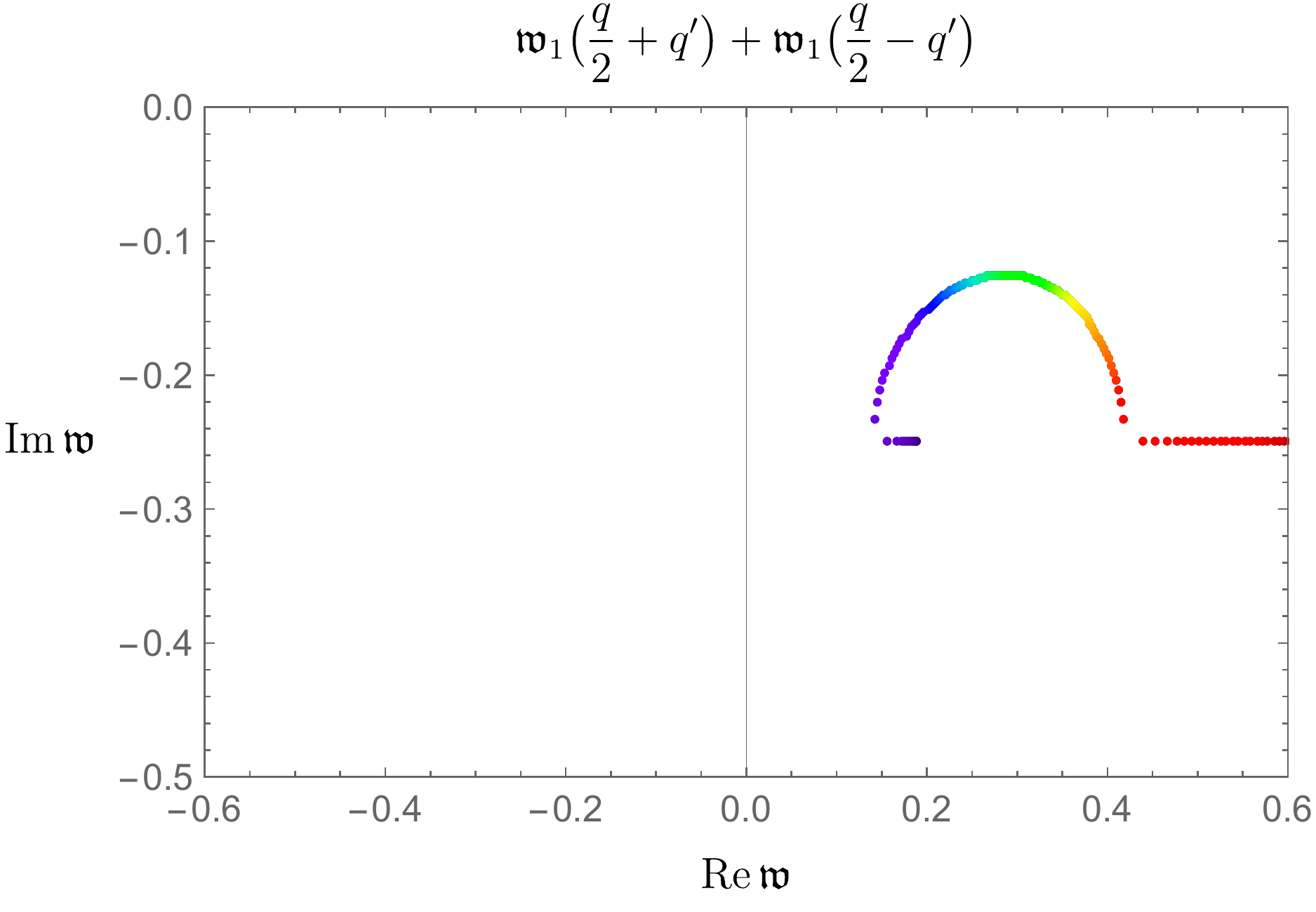}\,\,\,\,\includegraphics[width=0.48\textwidth]{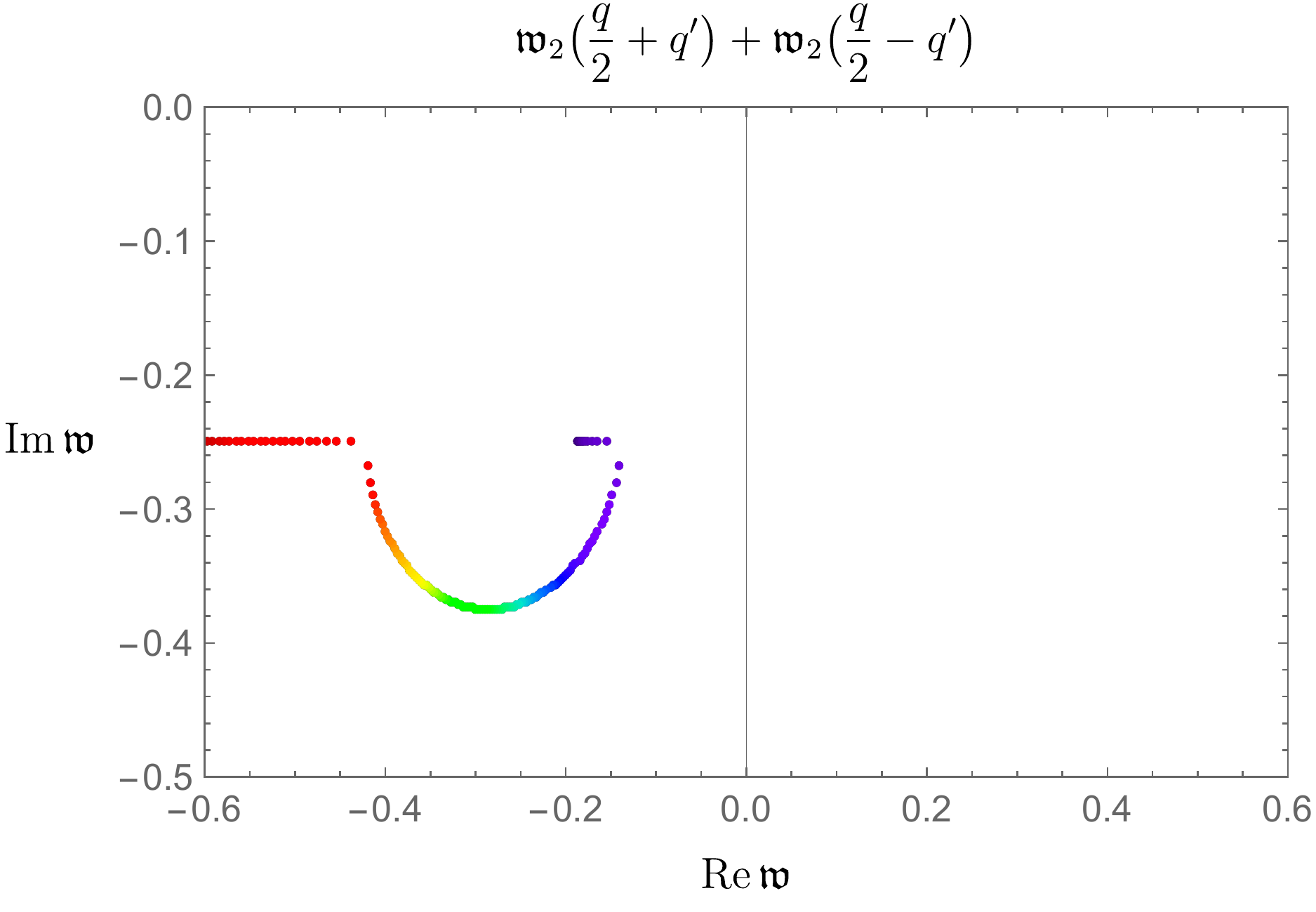}\,
		\includegraphics[width=0.48\textwidth]{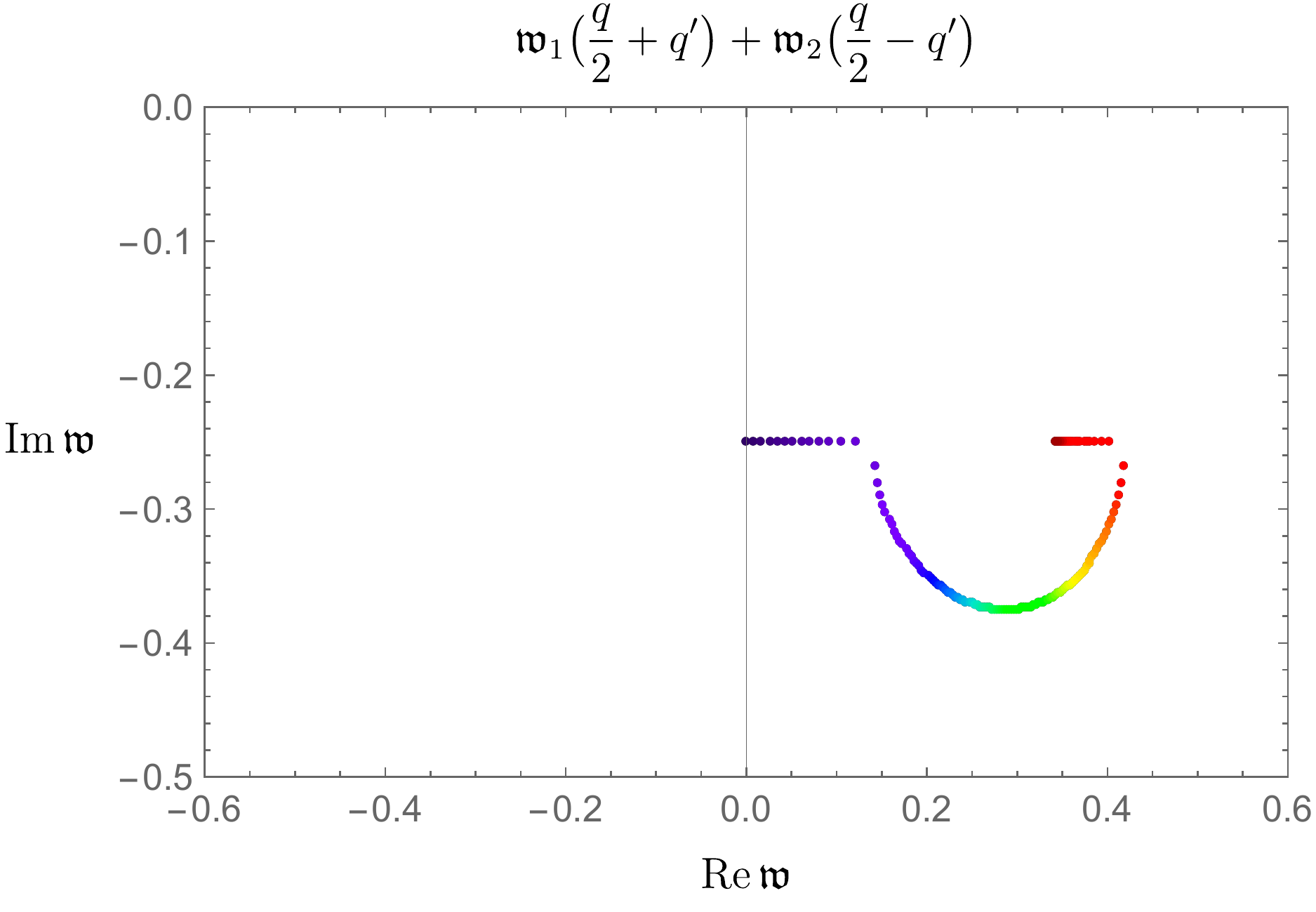}\,\,\,\,\includegraphics[width=0.48\textwidth]{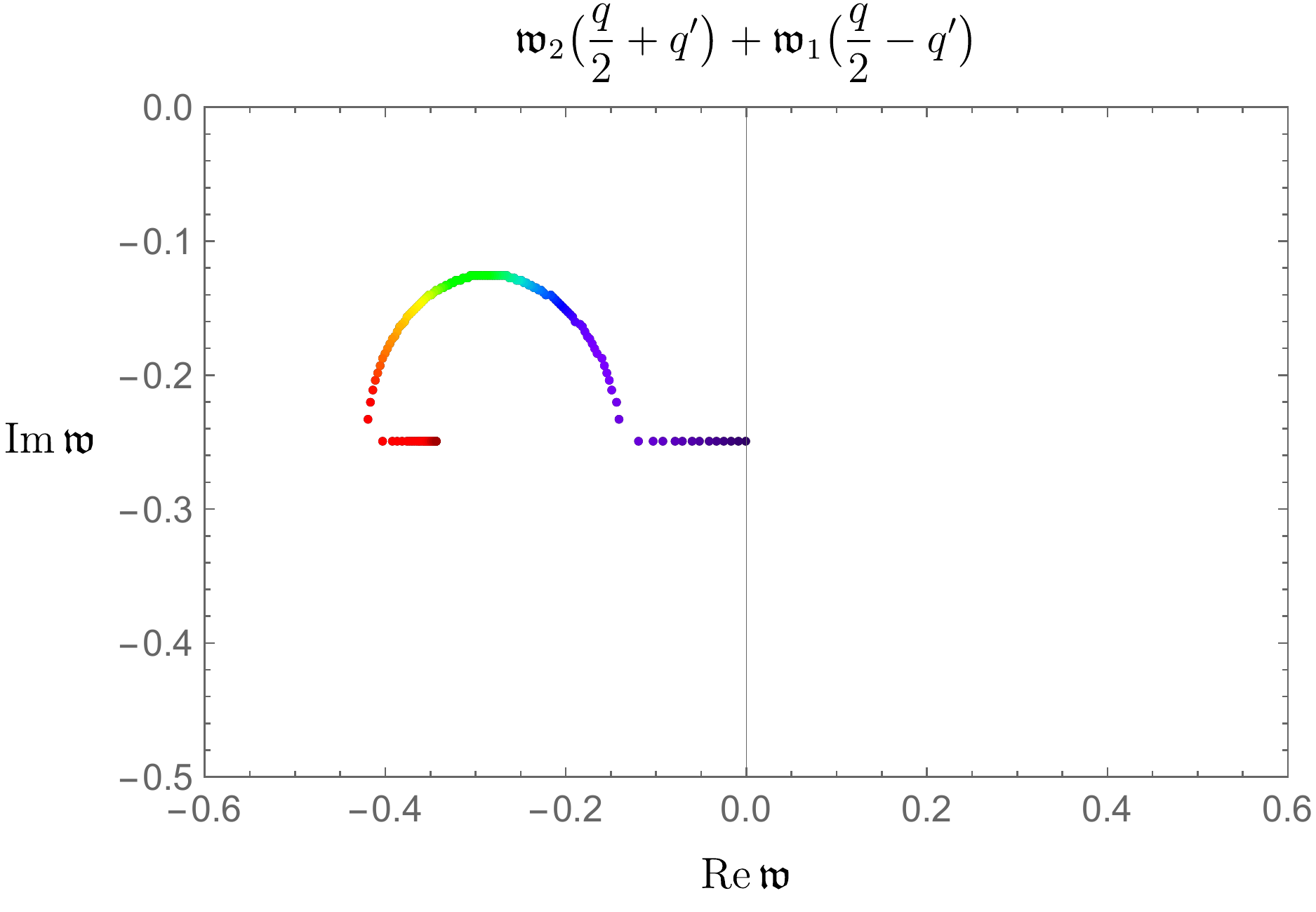}\,
		\caption{Illustration of the $\mathcal{S}^{ij}_{\qn}$'s at $\qn=(0,0,q=1.25)$ in the lower half of the complex frequency plane. In each panel, we have shown values of $\wn_i\big(\frac{q}{2}+q'\big)+\wn_j\big(\frac{q}{2}-q'\big)$ for 150 different $q'$ values between $0\le q'<1.5$. }
		\label{four_parts_2nd}
	\end{figure}

	By increasing the value of $\qn$, $\wn_{11}$ and $\wn_{22}$ move closer to each other and collide at $\qn=1$. As the second case to illustrate, we have considered $\qn=1.25$. Repeating the procedure used at $\qn=0.5$, we solve equations \eqref{conditions} numerically, by use of \eqref{S_qn_aplplied}. The four $\mathcal{S}_{\qn}^{ij}$ are shown in figure \ref{four_parts_2nd}. The entire branch cut structure and threshold singularities are given in the bottom left panel of figure \ref{Compare_before}. Again, we see that the results are exactly in line with the most recent results of ref. \cite{Abbasi:2022aao} in the theory of UV-regulated non-linear diffusion, shown in the bottom right panel.
		\vspace{2mm}
	
	Let us emphasize that the left and right panels of figure \ref{Compare_before} have been found through two different methods; the left figures were obtained by developing a new numerical method to solve the on-shell equations. However, the right panel results are the result of explicit loop calculations in ref. \cite{Abbasi:2022aao}. The branch-cut structure shown in the left panel of Figure \ref{Compare_before} is the central result of this section.
\subsubsection{Solving Landau equations}
	\label{threshold}
The on-shell conditions \eqref{conditions} identify where singularities can occur in the integrand. However, it is not sufficient for these singularities to develop in the full integral. Because we can deform the integral contour in a way that avoids these singularities. The singularity of the integrand will be the singularity of the integral, i.e.; a branch point, if it cannot be avoided by doing any contour deformation. One way to obtain these points is that the integration contour is pinched between the singularities of the integrand. Therefore, such points are called``pinch singularities"   (see Appendix for details).
		\vspace{2mm}

Landau developed a method of finding pinched singularities (among other singularities) of integrand in perturbation quantum field theory. To briefly see how, let us rewrite the denominator of the integrand in \eqref{Feynman_parameter} as
\begin{equation}\label{f_Landau}
	f^2\equiv\big[\alpha_1\, \mathcal{D}(\wn'\,,\qn')+\alpha_2\, \mathcal{D}(\wn-\wn'\,,\qn-\qn')\big]^2\,.
\end{equation}
We can always make a transformation in the variables $\wn'$ and $\qn'$ to eliminate the terms linear with respect to the integral variables: $\wn'\equiv\wn'(\wn'')$ and $\qn'\equiv\qn'(\qn'')$ where $\wn'$ (and $\qn'$) differs from $\wn''$ (and $\qn''$) by a constant. It is easy to find the explicit form of the transformation, however, we only need the final form of $f$ 
\begin{equation}\label{f_phi_K}
f=\varphi(\wn, \qn;\alpha_i) + K(\wn''^{\,2},\qn''^{\,2}; \alpha_i)
\end{equation}
where
\begin{eqnarray}\label{varphi}
\varphi(\wn, \qn;\alpha_i)	&=&\frac{4 (\alpha_1^2+\alpha_2^2)}{\theta_{\eta} (\alpha_1+\alpha_2)}+\frac{\alpha_1\,\alpha_2}{\theta (\alpha_1+\alpha_2)}\big(\theta_{\eta}^2 \wn^2+8 i \theta_{\eta} \wn-\theta_{\eta} \qn^2-8\big)\\
K(\wn''^2,\qn''^2)&=&(\alpha_1+\alpha_2)\big(\theta_{\eta} \wn''^{\,2}-\qn''^{\,2}\big)
\end{eqnarray}
 We can make another change of variable $q''\rightarrow i \tilde{\qn}''$ to make $K$ positive definite: 
\begin{equation}\label{}
K(\wn''^{\,2},\qn''^{\,2};;\alpha_i)=(\alpha_1+\alpha_2)\,\tau\, \wn''^{\,2}+\tilde{\qn}''^{\,2}
\end{equation}
Now if $\varphi>0$ for all values of $\alpha_i$, performing $\wn''-$ and $\tilde{\qn}''-$integration do not develop any singularity. On the other hand if for some values of $\alpha_i$ we have $\varphi<0$, the integral becomes complex. In this case,  for a given momentum $\qn$, the singularity of integral corresponds to values of $\wn(\qn)$ where $\varphi$ vanishes for some particular values of $\alpha_i$ and is positive at all other values of $\alpha_i$. In other words by treating $\varphi$ as being a function of $\alpha_i$, the singularities corresponds to vanishing of $\varphi(\alpha_i)$ at its extremum point:
\begin{equation}\label{extremum}
\varphi(\alpha_i)=0\,,\,\,\,\,\,\,\,\,\,	\frac{\partial \varphi(\alpha_i)}{\partial \alpha_i}=0\,.
\end{equation}
To simplify the second condition above, let us denote that from \eqref{f_phi_K} it is clear that
\begin{equation}\label{}
	\varphi(\alpha_i)=f\bigg|_{\frac{\partial f}{\partial \wn''}=0, \,\frac{\partial f}{\partial \qn''}=0}\,.
\end{equation}
Since $\wn'$ (and $\qn'$) differs from $\wn''$ (and $\qn''$) by a constant, the above two conditions can also be written as 
\begin{equation}\label{lanndau_loop}\boxed{
\frac{\partial f}{\partial \wn'}=0\,,\,\,\,\,\,\,\, \,\frac{\partial f}{\partial \qn'}=0}
\end{equation}
Therefore $\varphi(\alpha_i)$ is the same as $f$ when the above additional conditions are imposed. Thus instead of differentiating $\varphi$ in the second equation in \eqref{extremum},  we can differentiate $f$ with respect to $\alpha_i$ and then impose the two conditions above. We arrive at 
\begin{equation}\label{}
	\frac{\partial \varphi(\alpha_i)}{\partial \alpha_i}=\,\frac{\partial f}{\partial \alpha_i}=\,0
\end{equation}
Applying this to \eqref{f_Landau}, we get the two on-shell conditions
\begin{equation}\label{landao_on_shell}\boxed{
\mathcal{D}(\wn'\,,\qn')=0\,,\,\,\,\,\,\,\,\,\, \mathcal{D}(\wn-\wn'\,,\qn-\qn')=\,0}
\end{equation}
Equations \eqref{lanndau_loop} and \eqref{landao_on_shell} are the Landau conditions/equations for the BDNK theory, at 1-loop order. Equations \eqref{lanndau_loop} are also called ``\textit{Landau loop equations}". In the Appendix, we review the original equations derived by Landau for the scalar field theory. Compared to the on-shell condition in the original derivation of Landau (see \eqref{Landau_on_shell}), here our on-shell condition is not a covariant equation (see  \eqref{on_shell}). 
		\vspace{2mm}
		
	Before solving the Landau equation, let us express that any point $(\wn, \qn)$ satisfying \eqref{lanndau_loop} and \eqref{landao_on_shell} is indeed a pinched singularity of the integrand. We need to remember that when two curves are tangent, their normal vectors at the point of intersection are parallel. In other words, a linear combination of normal vectors vanishes. The equations \eqref{lanndau_loop} are very similar to this case; they can be written as
	\begin{equation}\label{lanndau_loop_2}
		\begin{split}
		\alpha_1\, \frac{\partial \mathcal{D}(\wn'\,,\qn')}{\partial\wn'}+\alpha_2\, \frac{\partial\mathcal{D}(\wn-\wn'\,,\qn-\qn')}{\partial\wn'}&=\,0\\
		\alpha_1\, \frac{\partial \mathcal{D}(\wn'\,,\qn')}{\partial\qn'}+\alpha_2\, \frac{\partial\mathcal{D}(\wn-\wn'\,,\qn-\qn')}{\partial\qn'}&=\,0
	\end{split}
\end{equation}
We see that the gradient of on-shell conditions (with respect to loop momentum and frequency) is linearly dependent.
This shows that solving \eqref{landao_on_shell} and \eqref{lanndau_loop_2} is equivalent to finding pinch singularities of the integrand on the integral contour.
				\vspace{2mm}
				
	Let us now proceed to solve the Landau loop equations:
	\begin{equation}\label{}
		\begin{split}
			\frac{\partial f}{\partial \wn'}=0&\,\,\,\,\,\rightarrow\,\,\,\,\,\,\wn'=\frac{\alpha_2}{\alpha_1+\alpha_2}\wn+\frac{\alpha_2-\alpha_1}{\alpha_2+\alpha_1}\frac{2 i }{\theta_{\eta}}\,,\\
			\frac{\partial f}{\partial \qn'}=0&\,\,\,\,\,\rightarrow\,\,\,\,\,\,\qn'=\frac{\alpha_2}{\alpha_1+\alpha_2}\qn\,.
		\end{split}
	\end{equation}
	Substituting the above two expressions into the on-shell conditions, we arrive at
	\begin{subequations}\label{}
		\begin{align}\label{on_shell_1}
		\frac{\partial f}{\partial \alpha_1}=0&\,\,\,\,\,\rightarrow\,\,\,\,\,\,0=4-\frac{\alpha_1^2}{(\alpha_1+\alpha_2)^2}\,\big(\theta_{\eta}\qn^2+(4- i \theta_{\eta} \wn)^2\big)\,,\\\label{on_shell_2}
			\frac{\partial f}{\partial \alpha_2}=0&\,\,\,\,\,\rightarrow\,\,\,\,\,\,0=4-\frac{\alpha_2^2}{(\alpha_1+\alpha_2)^2}\,\big(\theta_{\eta}\qn^2+(4- i \theta_{\eta} \wn)^2\big)\,.
		\end{align}
	\end{subequations}
Now we  perform the integration over $\alpha_2$, which replaces $\alpha_2$ with $1-\alpha_1$. We also omit the subscript of $\alpha_1$.	
	From \eqref{on_shell_1}, we find the solution for $\alpha$, i.e., $\alpha^*$, as 
	\begin{equation}\label{Landau_sol}
		\alpha^*=\frac{2}{\sqrt{\theta_{\eta} \qn^2+(4-i \theta_{\eta} \wn)^2}}\,.
	\end{equation}
	According to ref. \cite{Landau:1959}, the above $\alpha^*$ may be associated with the threshold singularity of the response function. Substituting this value into the second on-shell condition \eqref{on_shell_2}, we arrive at 
	\begin{equation}\label{}
		-16-\theta_{\eta} \qn^2+ 8 i \theta_{\eta} \wn+\theta_{\eta}^2 \wn^2+4 \sqrt{\theta_{\eta} \qn^2+(4-i \theta_{\eta} \wn)^2}=\,0\,.
	\end{equation}
	This equation \footnote{This kind of equation that constrains  the external momenta and frequencies to obey the Landau equations is called the ``Landau curve'' \cite{Fairlie}.} has four roots: 
	\begin{subequations}\label{}
		\begin{align}\label{threshold_1}
			\wn=&-\frac{i}{\theta_{\eta}}\left(4\pm\sqrt{16-\theta_{\eta} \qn^2}\right)\,,\\\label{threshold_2}
			\wn=&-\frac{4i}{\theta_{\eta}}\pm|\qn|\sqrt{\frac{1}{\theta_{\eta}}}\,.
		\end{align}
	\end{subequations}
	It is easy to check that the two roots identified by \eqref{threshold_1} correspond to the threshold singularities $\wn_{11}$ and $\wn_{22}$ in figure \ref{Compare_before}, while the roots given by \eqref{threshold_2} correspond to  $\wn_{12}$ and $\wn_{21}$ in the same figure. Note that  $\wn_{ij}$'s shown in figure  \ref{Compare_before} were all found numerically; now, here we see that they can be expressed by two explicit analytic formulas \eqref{threshold_1} and \eqref{threshold_2}.
		\vspace{2mm}
		
	Let us now evaluate \eqref{Landau_sol} at \eqref{threshold_1} and \eqref{threshold_2}. One finds
	\begin{subequations}\label{}
		\begin{align}\label{alpha_star_1}
			\alpha_1=\,\alpha_2\,\,\,\equiv\,\,	\alpha^*\big|_{\text{at}\,\eqref{threshold_1}}\,\,\,\,\,=&\,\,\,\,\frac{1}{2}\\\label{alpha_star_2}
			\alpha_1=1-\alpha_2\,\equiv\,\,	\alpha^*\big|_{\text{at}\,\eqref{threshold_2}}\,\,\,\,\rightarrow&\,+\infty
		\end{align}
	\end{subequations}
	This simply tells us that $\wn_{11}$ and $\wn_{22}$ are normal thresholds, i.e., $0<\alpha^* <1$ \cite{Hannesdottir:2022xki},  while $\wn_{12}$ and $\wn_{21}$ are second-type singularities, because $\alpha^*$ is infinite \cite{Fairlie} (see Appendix \ref{Landau_App} for more details on these two types of singularities).
		\vspace{2mm}
	
	Let us summarize. We first numerically specified the branch cut structure of the shear stress response function, based on the formula \eqref{S_qn_aplplied}. The results are given in the left panel of the figure \ref{Compare_before}. Then to find the threshold singularities shown in the figure, i.e., $\wn_{i,j}: i,j\in\{1,2\}$, we analytically solved the well-known Landau equations. \textbf{The four threshold singularities found in \eqref{threshold_1} and \eqref{threshold_2}, and the branch cut structure of the figure \ref{Compare_before} are exactly consistent with the results of the loop calculations in ref. \cite{Abbasi:2022aao}}. 
		\vspace{2mm}
		
	As discussed at the beginning of \sec{reason}, knowledge of the analytic structure  is useful for discovering discontinuities in scattering amplitudes in field theory. The latter can then be used to find some specific decay rates or cross sections. These field theory quantities are also calculated in EFT of hydrodynamics \cite{Endlich:2010hf,Gripaios:2014yha}.
	However, to the best of our knowledge, so far, no work has used the analytic structure discussed in our paper to find such quantities in the context of hydrodynamics. Our work is actually a first step in this direction. We leave the explicit field theory calculations to future work.

	\subsection{Long-time tails}
	\label{long_time}
	In previous sections we elaborated on the analytic structure of $G^{R}_{T_{xy}T_{xy}}(\omega,\textbf{k})$. The result is simple. In the linear regime it is analytic in  the entire lower half complex $\omega$ plane except for the location of two simple poles \eqref{on_shell_D_T}. However, we showed that nonlinear effects significantly affects this behavior. For example in the small $\qn$ limit $G^{R}_{T_{xy}T_{xy}}(\omega,\textbf{k})$ found the complicated analytic structure illustrated in the top panel of figure \ref{Compare_before}.
		   		\vspace{2mm}
		   		
		Our man goal in this section is to investigate the effect of nonlinearities, in particular the analytic structure shown in  figure \ref{Compare_before},  on the late time behavior of  $G^{R}_{T_{xy}T_{xy}}(t,\textbf{k})$ defined as following
	\begin{equation}\label{G_R_t}
		G^{R}_{T_{xy}T_{xy}}(t,\textbf{k})=\,\int\frac{d\omega}{2\pi} G^{R}_{T_{xy}T_{xy}}(\omega,\textbf{k}) e^{- i\omega t}\,.
	\end{equation}
We aim to do this without explicitly evaluating the above Fourier integral. The reason is simply that we  do not access the closed form of $G^{R}_{T_{xy}T_{xy}}(\omega,\textbf{k})$; we only have some information about its analytic properties. However we will show that this  information is sufficient to find the behavior of $G^{R}_{T_{xy}T_{xy}}(t,\textbf{k})$ at late times. 
			   		\vspace{2mm}
	
	To proceed, let us mention a few comments which will guide us to track the path.
	\begin{itemize}		   		
		\item Nonlinear effects cause long-time tails in correlation functions. This behavior is directly related to the presence of singularities in the momentum space correlation functions \cite{Kovtun:2003vj}.
		\item  In BDNk theory, the shear stress correlation function has four branch point singularities (see top panel of figure \ref{Compare_before}). One would expect that at late times, i.e., times larger than any specific time-scale of the theory, the branch point $\wn_{11}$ to give the dominant contribution. It can be simply understood by looking at $e^{- i \omega t}$ in the integrand; this factor is exponentially decaying in the vicinity of each of the branch points. Clearly, the decay associated with $\wn_{12}$, $\wn_{21}$ and $\wn_{22}$ is much faster than that of $\wn_{11}$.
		This justifies that at late times,  $G^{R}_{T_{xy}T_{xy}}(\omega,\textbf{k})$ should be replaced by its leading order singularity around $\wn_{11}$. 
		\item \textit{How to find the leading order singularity around a branch point}? 
			This question has been also answered  in the original paper by Landau \cite{Landau:1959}. 
			\end{itemize}		   		
			  Thus, what we are going to do is to first introduce  Landau's method of finding leading singularity of the integral in the vicionithy of a branch point. Then we will apply it to $G^{R}_{T_{xy}T_{xy}}(\omega,\textbf{k})$ of the BDNK theory near the branch point $\wn_{11}$. Finally, we will evaluate \eqref{G_R_t} to find the long-time tail.
			   		\vspace{2mm}
\subsubsection{Nature of singularities }
\label{nature}
		Let us assume that the extremum value of $\varphi(\wn, \qn;\alpha_i)$ in \eqref{f_phi_K} is $\varphi(\wn, \qn; \alpha^*)$. It was shown by Landau that the leading singularity of the Integral is then given by \cite{Landau:1959}
	\begin{equation}\label{}
		\sim  \text{constant}\,\cdot\,\big( \varphi(\wn, \qn;\alpha^*)\big)^{\frac{1}{2}m-n}
	\end{equation}
	   where $m$ is the number of independent integrations and $n$ is the number of internal lines in the corresponding Feynman integral. In Appendix, we have evaluated $m$ in $d-$dimension in terms of $n$ and $V$  ($V$ is the number vertices in the diagram).  For the simple bubble diagram corresponding to \eqref{Feynman_parameter}, clearly $n=2$ and $m=4+1$. In the latter, $4$ refers to integration over $\wn$ and $\qn$, while $1$ corresponds to integration over one single independent $\alpha$. As a result, near the threshold  specified by the solution of Landau equations, the integral corresponding to the bubble diagram behaves as \footnote{The constant in front of \eqref{nature_G_R} can be found by using the discussion given in the end of section (1.5) in \cite{Eden:1966}.}
	   \begin{equation}\label{nature_G_R}
G^{R}_{T_{xy}T_{xy}}(\wn,\qn)\equiv \mathcal{I}(\wn, \qn)\sim \text{constant}\cdot\bigg(\varphi(\wn, \qn;\alpha^*)\bigg)^{\frac{1}{2}}\,.
	   \end{equation}
	   In next subsection we will evaluate this expression in BDNK theory and then use the result to find the long-time tail behavior of the $G^{R}_{T_{xy}T_{xy}}(t,\textbf{k})$.
	\subsubsection{Long-time tails in BDNK theory}
	\label{long_tail_calcualtion}
	  The expression of $\varphi(\wn, \qn; \alpha)$ for the shear stress in BDNK theory was already found in \eqref{varphi}. However, as discussed before, $\mathcal{I}(\wn,\qn)$ has four branch point singularities $\wn_{ij}:\, i,j\in\{1,2\}$.  The value of $\alpha^*$ corresponding to $\wn_{11}$ is $1/2$ (see \eqref{alpha_star_1} and \eqref{alpha_star_2}).
Using the latter, we find
\begin{equation}\label{phi_star}
\varphi\big(\omega, \textbf{k}; \alpha^*=1/2\big)=\,\omega+i\frac{\eta}{2w}\textbf{k}^2-i \frac{\theta}{2w}\omega^2\,.
	   \end{equation}
	Note that we have used \eqref{re_scale} to return all quantities to their dimensionful versions ($w=4p$).  Since now on, we will represent the results in terms of dimensionful quantities.
				   		\vspace{2mm}
	
	Let us emphasize that $\alpha^*=1/2$ corresponds to $\omega_{22}$ as well. In other words, setting $\varphi\big(\omega, \textbf{k}; \alpha^*=1/2\big)=0$ in \eqref{phi_star} gives the two branch points $\omega_{11}$ and $\omega_{22}$. But we only need to find the leading singularity near $\omega_{11}$.  For this reason, we change the variable as $\omega=\omega_{11}+ i\bar{\omega}$, and expand $\varphi\big(\omega, \textbf{k}; \alpha^*=1/2\big)$ about $\bar{\omega}=0$ to first non-vanishing order. We find:
	\begin{equation}\label{phi_star_expand}
		\varphi\big(\bar{\omega}, \textbf{k}; \alpha^*=1/2\big)=\,\sqrt{1-\theta \eta\big(\frac{\textbf{k}}{w}\big)^2}\,\,\bar{\omega}\,+\mathcal{O}\big(\bar{\omega}^2\big)
	\end{equation}
	This through  \eqref{nature_G_R} specifies the leading singularity of the    $G^{R}_{T_{xy}T_{xy}}(\omega,\textbf{k})$ near $\omega_{11}$. Substituting into  \eqref{G_R_t}, we then arrive at \footnote{The pre-factor $\frac{w^2}{\eta^{3/2}}$ in \eqref{G_R_t_subst} has two parts; $w^2$ comes from \eqref{int_plus_plus} while $1/\eta^{3/2}$ comes from the ``constant" sitting in front of \eqref{nature_G_R}.}
	\begin{equation}\label{G_R_t_subst}
		G^{R}_{T_{xy}T_{xy}}(t,\textbf{k})\sim \,\frac{w^2}{\eta^{3/2}}\left[1-\theta\eta\big(\frac{\textbf{k}}{w}\big)^2\right]^{\frac{1}{4}}e^{- i \big(1-\sqrt{1-\theta\eta(\frac{\textbf{k}}{w})^2}\big) \frac{w}{\theta}t }\int_{+i \infty+ i \omega_{11}}^{-i \infty+ i \omega_{11}}\frac{id\bar{\omega}}{2\pi} \,\sqrt{\bar{\omega}}\,e^{\omega t}\,.
	\end{equation}
	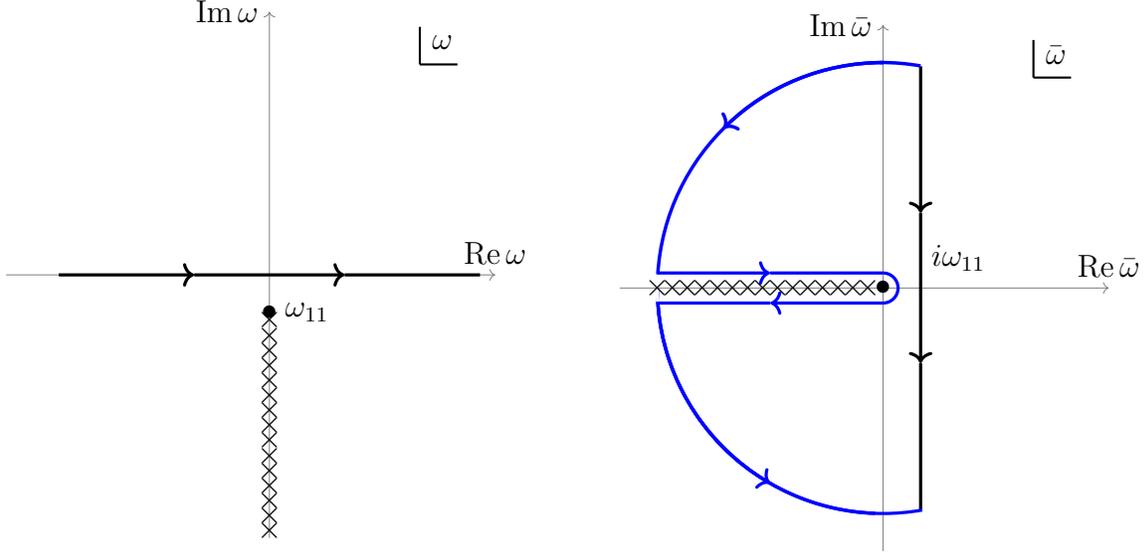
\begin{figure}
		\centering
			\begin{tikzpicture}
			[decoration={markings,
			}
			]
			\draw[help lines,->] (-3.5,0) -- (3.,0) coordinate (xaxis);
			\draw[help lines,->] (0,-3.5) -- (0,3.5) coordinate (yaxis);
			\draw[draw,black,line width=1.2pt,->] (-2.8,0) -- (-1,0) coordinate;
			\draw[draw,black,line width=1.2pt,->] (-1,0) -- (1,0) coordinate;
			\draw[draw,black,line width=1.2pt] (1,0) -- (2.8,0) coordinate;
						\draw[draw,black,line width=.8pt] (2,2.8) -- (2.5,2.8) coordinate;
												\draw[draw,black,line width=.8pt] (2,2.8) -- (2,3.3) coordinate;
			
						\node[] at (0,-3.4) {$\times$};
						\node[] at (0,-3.2) {$\times$};
			\node[] at (0,-3) {$\times$};
			\node[] at (0,-2.8) {$\times$};
			\node[] at (0,-2.6) {$\times$};
			\node[] at (0,-2.4) {$\times$};
			\node[] at (0,-2.2) {$\times$};
			\node[] at (0,-2.) {$\times$};
			\node[] at (0,-1.8) {$\times$};	
			\node[] at (0,-1.6) {$\times$};	
			\node[] at (0,-1.4) {$\times$};
			\node[] at (0,-1.2) {$\times$};
			\node[] at (0,-1.) {$\times$};
			\node[] at (0,-.8) {$\times$};	
						\node[] at (0,-.6) {$\times$};			
			\node[] at (0,-.5) {$\bullet$};
			\node[above] at (xaxis) {$\text{Re}\, \omega$};
			\node[left] at (yaxis) {$\text{Im} \,\omega$};
			\node[black] at (.5,-.5) {$\omega_{11}$};
						\node[black] at (2.3,3.1) {$\omega$};
		\end{tikzpicture}
	\,\,\,\,\,\,\,\,\,\,\,\,
		\begin{tikzpicture}
			[decoration={markings,
			}
			]
			\draw[help lines,->] (-3.5,0) -- (3.,0) coordinate (xaxis);
			\draw[help lines,->] (0,-3.5) -- (0,3.5) coordinate (yaxis);
			\draw[draw,black,line width=1.2pt,->] (.5,2.95) -- (.5,1) coordinate;
						\draw[draw,black,line width=1.2pt,->] (.5,1) -- (.5,-1) coordinate;
									\draw[draw,black,line width=1.2pt] (.5,-1) -- (.5,-2.95) coordinate;
															\draw[draw,black,line width=.8pt] (2,2.8) -- (2.5,2.8) coordinate;
									\draw[draw,black,line width=.8pt] (2,2.8) -- (2,3.3) coordinate;
							\path[draw,line width=1.3pt,blue,postaction=decorate] (.5,2.955) arc (80:90.5:3);
												\path[draw,line width=1.3pt,blue,postaction=decorate,->] (0,3) arc (90:135:3);
					\path[draw,line width=1.3pt,blue,postaction=decorate] (0,3) arc (90:176.5:3);
			\path[draw,line width=1.3pt,blue,postaction=decorate] (-3,0) arc (180:268:3);
						\path[draw,line width=1.2pt,blue,postaction=decorate,->] (-3,0) arc (180:240:3);
			\path[draw,line width=1.2pt,blue,postaction=decorate] (-3,0) arc (180:280:3);
						\path[draw,line width=1.4pt,white,postaction=decorate] (-3,0) arc (180:183.5:3);
			\draw[draw,blue,line width=1.2pt,->] (0,-.2) -- (-1.5,-.2) coordinate;
			\draw[draw,blue,line width=1.2pt] (-3,-.2) -- (-1.5,-.2) coordinate;
				\draw[draw,blue,line width=1.2pt] (0,.2) -- (-1.5,.2) coordinate;
	\draw[draw,blue,line width=1.2pt,->] (-3,.2) -- (-1.5,.2) coordinate;
			\path[draw,line width=1.2pt,blue,postaction=decorate] (0,.2) arc (90:-90:.2);

						\node[] at (-3,0) {$\times$};
			\node[] at (-2.8,0) {$\times$};
			\node[] at (-2.6,0) {$\times$};
			\node[] at (-2.4,0) {$\times$};
			\node[] at (-2.2,0) {$\times$};
			\node[] at (-2.,0) {$\times$};
			\node[] at (-1.8,0) {$\times$};	
			\node[] at (-1.6,0) {$\times$};	
						\node[] at (-1.4,0) {$\times$};
			\node[] at (-1.2,0) {$\times$};
			\node[] at (-1.,0) {$\times$};
			\node[] at (-.8,0) {$\times$};	
			\node[] at (-.6,0) {$\times$};	
						\node[] at (-.4,0) {$\times$};
			\node[] at (-.2,0) {$\times$};	
			\node[] at (0,0) {$\bullet$};
			\node[above] at (xaxis) {$\text{Re}\, \bar{\omega}$};
			\node[left] at (yaxis) {$\text{Im} \,\bar{\omega}$};
			\node[black] at (1,.4) {$i \omega_{11}$};
					\node[black] at (2.3,3.1) {$\bar{\omega}$};
		\end{tikzpicture}
	\caption{ Left panel:  The dominant part of the analytic structure of $G^R(\omega, \textbf{k})$ at late times. Right pane: Changing the integral variable from $\omega$ to $\bar{\omega}$ and contour deformation. Note that $i \omega_{11}=\big(1-\sqrt{1-\theta \eta\frac{\textbf{k}^2}{w^2}}\big)\frac{w}{\theta}$ is real-valued.}
	\label{contour}
	\end{figure}
	The integration contour for this integral has been shown by black in the right panel of the figure \ref{contour}. We deform the contour across the branch cut of $\sqrt{\bar{\omega}}$ (gray in figure). Then we find	
	\begin{equation}\label{G_R_t_deform}
		G^{R}_{T_{xy}T_{xy}}(t,\textbf{k})\sim \,i\frac{w^2}{\eta^{3/2}}\left[1-\theta\eta\big(\frac{\textbf{k}}{w}\big)^2\right]^{\frac{1}{4}}e^{-  \big(1-\sqrt{1-\theta\eta(\frac{\textbf{k}}{w})^2}\big) \frac{w}{\theta}t }\int_{- \infty}^{0}\frac{d\bar{\omega}}{2\pi} \,\text{Disc}\sqrt{\bar{\omega}}\,e^{\omega t}\,.
	\end{equation}
	where $\text{Disc} f(z)=\lim\limits_{\epsilon\to 0}f(z+ i \epsilon)-f(z-i \epsilon)$. Performing the above integral we find:
	\begin{equation}\label{G_R}\boxed{
		G^{R}_{T_{xy}T_{xy}}(t,\textbf{k})\sim \,w^{1/2}\left[1-\theta\eta\big(\frac{\textbf{k}}{w}\big)^2\right]^{\frac{1}{4}}\frac{e^{-  \big(1-\sqrt{1-\theta\eta(\frac{\textbf{k}}{w})^2}\big) \frac{w}{\theta}t }}{(\gamma_{\eta} t)^{3/2}}}
	\end{equation}
with $\gamma_{\eta}=\eta/w$. This is actually the central result of this subsection. It should be note that this is a late time result; it can only  be used to describe the decay of correlation function at times larger than the ``diffusion time". As mentioned earlier, the rate of transverse momentum diffusion is given by the expression in front of $t$ in the exponential, that we call it  $\Gamma_{D}$. Then the diffusion time is defined as  
\begin{equation}\label{diffuion_time}
t_{D}\equiv\Gamma_{D}^{-1}=\, \bigg[\big(1-\sqrt{1-\theta\eta(\frac{\textbf{k}}{w})^2}\big) \frac{w}{\theta}\bigg]^{-1}
\end{equation}
 It is worth emphasizing that in the conventional hydrodynamics, $\theta=0$. In this case the above equation becomes
\begin{equation}\label{theta_set_0}
	G^{R}_{T_{xy}T_{xy}}(t,\textbf{k})\sim \,w^{1/2}\frac{e^{-  \frac{1}{2}\gamma_{\eta} \textbf{k}^2 t }}{(\gamma_{\eta} t)^{3/2}}\,.
\end{equation}
As one expects, there is a  tail with fractional power  together with an exponential factor \cite{Delacretaz:2020nit}. The decay  of the  exponential term is controlled by $\frac{1}{2}\gamma_{\eta}\textbf{k}^2$ which is familiar from conventional hydro \cite{Delacretaz:2020nit}. In another familiar limit, we can reproduce the well-known result of \cite{Kovtun:2003vj} at zero momentum:
\begin{equation}\label{}
	G^{R}_{T_{xy}T_{xy}}(t,\textbf{k}=0)\sim \,\frac{w^{1/2}}{(\gamma_{\eta} t)^{3/2}}\,.
\end{equation}
This is the same as equation (42) in the mentioned reference.
				   		\vspace{2mm}
				   		
				   		\begin{figure}[tb]
				   			\centering
				   			\centering
				   			\includegraphics[width=0.5\textwidth]{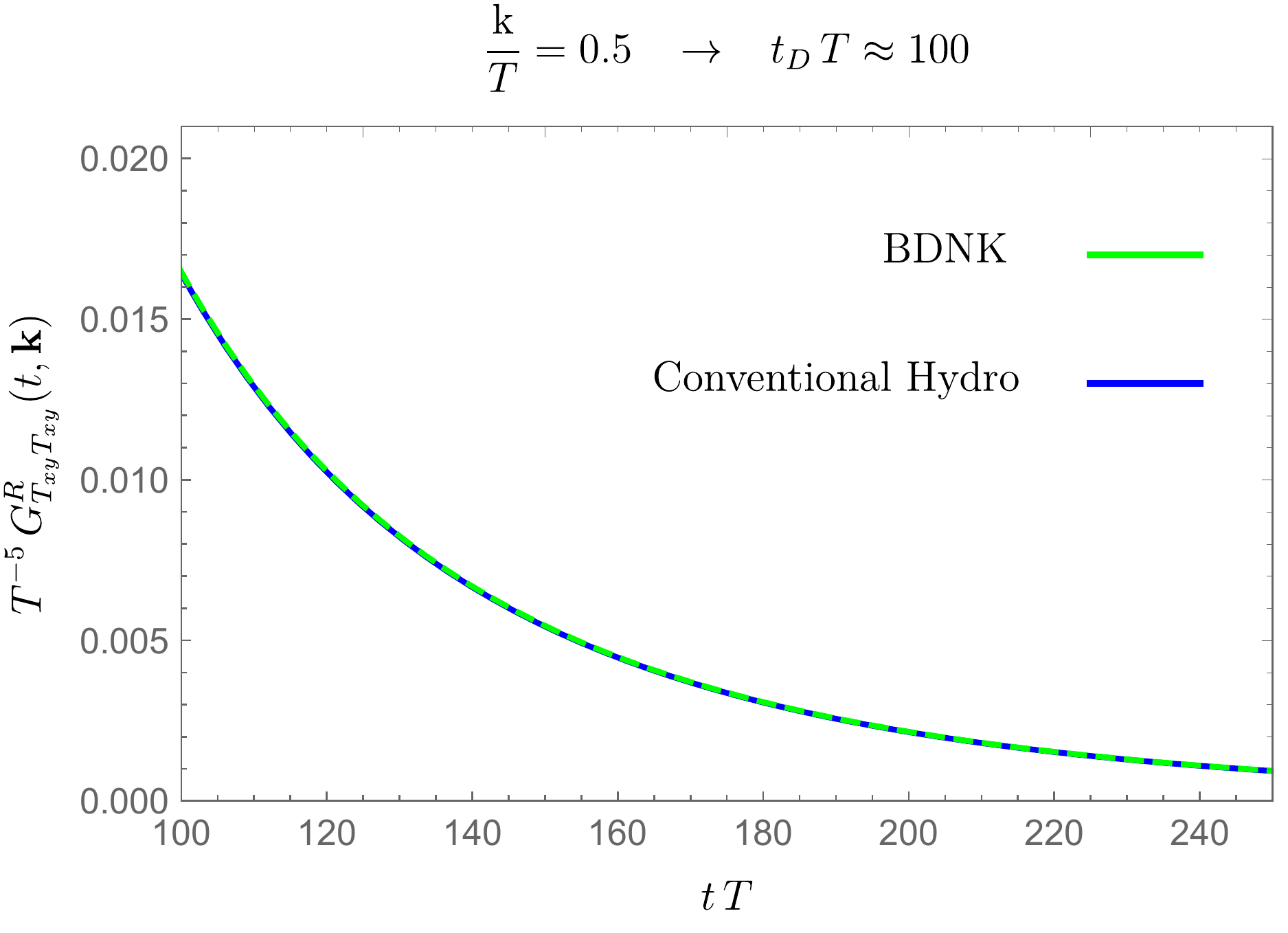}\,\,\,\,\includegraphics[width=0.48\textwidth]{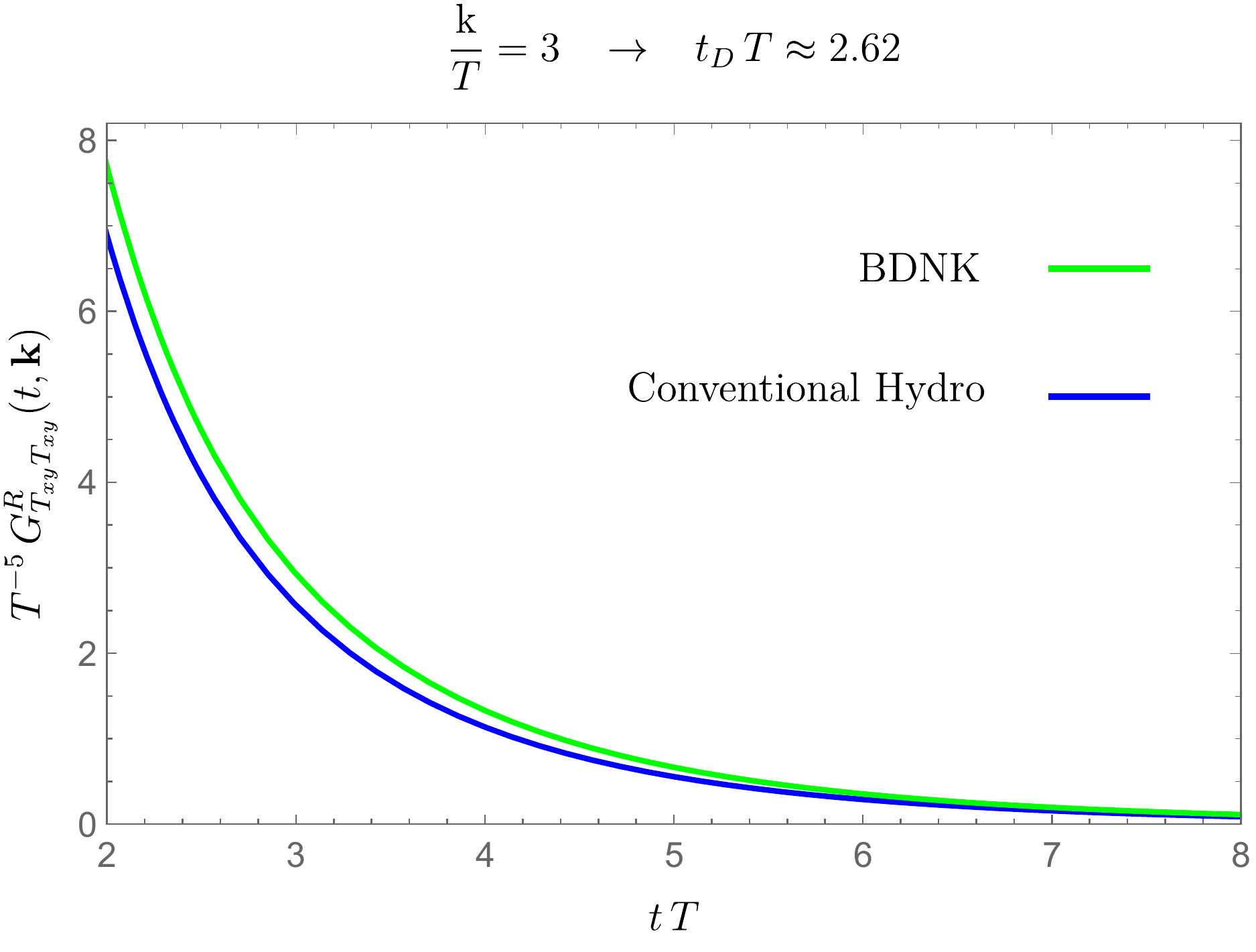}\,
				   			\caption{Long-time tail comparison between BDNK and the conventional first order relativistic hydrodynamics for $\frac{\theta}{s}= \frac{4\eta}{s}= \frac{1}{\pi}$. Both panels have been illustrated for times larger than the diffusion time of the transverse momentum, i.e.; $t\gtrsim t_D$.  Left panel: For relatively small momenta, nonlinear effects in BDNK theory are consistent with the same effects in  the conventional hydro. Right panel: At large momentum, the nonlinear effects in BDNK theory lead to slower decay of the response function than those in conventional hydro. }
				   			\label{long_time_tail}
				   		\end{figure}
				   		
				   		In order to gain more insight about the difference between BDNK theory and conventional hydrodynamics, we proceed with illustrating behavior of $G^{R}_{T_{xy}T_{xy}}(t,\textbf{k})$ for both cases in figure \ref{long_time_tail}. In order to explain how we have chosen the values of $\text{k}$ in the figure, let us recall the dispersion relations in the linear regime \eqref{on_shell_D_T}. As mentioned around figure.~\ref{pole_sound}, for $\theta_{\eta}=4$, the range of validity of BDNK theory is $\qn\lesssim 0.5$. In terms of dimensionfull quantities, it is written as (see \eqref{re_scale})
\begin{equation}\label{k_constartint}
\frac{\text{k}}{T}\,\lesssim\,\frac{1}{8}\bigg(\frac{\eta}{s}\bigg)^{-1}\,.
\end{equation}
Motivated by the value of $\eta/s$ frequently used for quark gluon plasma, we set $\frac{\eta}{s}= \frac{1}{4\pi}$. Then \eqref{k_constartint} simply tells us that we can use momenta within the range $\frac{\text{k}}{T }\lesssim\frac{\pi}{2}$. For illustrative purposes, we have taken two different values for $\frac{\text{k}}{T}$, one from this range and another from outside of this range; $\frac{\text{k}}{T}=0.5$ and $3$. The corresponding diffusion times can be found from \eqref{diffuion_time}.  It turns out that for BDNK theory $t_{D}\,T\approx 100, \,\,2.62 $, respectively. For conventional hydro, with $\theta=0$, the corresponding values are found to be very close to these values. So we take them the same values of $t_{D}$, namely $t_{D}\,T\approx 100, \,\,2.62 $,  for the two corresponding conventional hydro cases. Then for each choice of $\text{k}$, we show the decay of correlation function within the range $t\gtrsim t_{D}$.  
				   		\vspace{2mm}
				   		
				   		As it is seen in the figure,
				   		\begin{itemize}
				   			\item At small value of $\frac{\text{k}}{T}$, the BDNK results are not indistinguishable from the long-time tail caused by nonlinear effects in conventional hydrodynamics. Although the same result for the linear effects at small momentum could be predicted, however, the current result shows that at small momenta, even nonlinearities are not sensitive to UV physics.
				   			 This is another way of saying that UV-regulators in BDNK theory are irrelevant in the low IR limit.  This can also be clearly understood when taking the small momentum limit in \eqref{G_R}
				   			\begin{equation}\label{}
				   				\lim_{\text{k}/T\rightarrow 0}G^{R}_{T_{xy}T_{xy}}(t,\textbf{k})\sim \,w^{1/2}\frac{e^{-  \frac{1}{2}\gamma_{\eta} \textbf{k}^2 t }}{(\gamma_{\eta} t)^{3/2}}\,.
				   			\end{equation}
				   			The result is the same as if $\theta$ were zero (see \eqref{theta_set_0}). We see that $\theta$ vanishes from the nonlinear results with small momentum.
				   			\item At momenta outside the range of validity of BDNK theory, the effect of UV-regulator is significant. As shown in the right panel of the figure, the decay of correlation function in BDNK theory decays more slowly than the correlation function in traditional hydro theory. This is of course due to the action of the UV $\theta$ regulator, which weakens diffusion of the transverse momentum. It can be understood by comparing $\omega_{11}$ within two theories: 
				   			\begin{eqnarray}\label{omega_11_hydro}
                             \text{Conventional Hydro:}&\,\,\,\,\,\,\,\,& \omega_{11}=\,-\frac{i}{2}\frac{\eta }{w}\textbf{k}^2\\
                                                          \text{BDNK:}&\,\,\,\,\,\,\,\,& \omega_{11}=\,-i\big(1-\sqrt{1-\theta\eta(\frac{\textbf{k}}{w})^2}\big) \frac{w}{\theta}
				   			\end{eqnarray}
				   			We see that $\theta$ pushes $\omega_{11}$ in \eqref{omega_11_hydro} away from the real axis in the lower half complex $\omega$ plane. It should be noted that this is just a mathematical result; the BDNK theory is not supposed to give reliable results at large momenta \footnote{We thank Pavel Kovtun for discussion on this point.}.
				   			\end{itemize}

	\section{Discussion and outlook}  
	\label{conclusion}
This work focuses on correlation functions in BDNK theory, especially in conformal systems. The first part of this work focuses on the derivation of momentum space correlation functions in the sound and shear channels.
The pole structure of the correlation function is analyzed and found to be consistent with the spectrum discovered by studying the linearized BDNK equations in \cite{Kovtun:2019hdm}. Our calculations also reveal a feature of the theory in which the correlation function of the energy density develops a range of negative values. It should be noted that this feature is not the case in the small momentum limit, where we expect the theory to be consistent with conventional hydrodynamics.
			\vspace{2mm}
			
In order to gain insight about nonlinear fluctuations in the theory we borrowed some methods from field theory to find the structure of the correlation function outside the linear response regime. By developing a numerical method for solving the on-shell conditions, we discovered the branch cut structure of the shear stress response function. We also solved the Landau conditions analytically to find the threshold singularities of the same response function. To the best of our knowledge, this is the first time these ideas have been introduced and applied in a fluid dynamics framework. It would be interesting to extend this analysis to the sound channel, where one would hope to see a richer structure of correlation functions.
		\vspace{2mm}
		
	To understand how the branch point singularities discovered above affect the real-space correlation function, we then investigated the late-time behavior of the stress tensor correlation function. We analyzed and calculated the long-time tail of this correlation function. Consistent with conventional hydrodynamics \cite{Kovtun:2003vj}, we found fractional power $\sim t^{-3/2}$ with an exponential decay factor. It turns out that in the small momentum limit, the exponential factor does not depend on the UV-regulator; in other words, the UV-regulator is irrelevant in the IR physics.			\vspace{2mm}
		
	Regarding the correlation functions in the nonlinear regime, it would be very interesting to construct an effective field theory corresponding to the stable first-order  hydrodynamics. In this way, interactions between hydrodynamic modes can be considered systematically. Some initial steps in this direction have recently been taken in ref.  \cite{Abbasi:2022aao}. The theory of UV-regulated nonlinear diffusion constructed in ref.  \cite{Abbasi:2022aao} can be regarded as the effective field theory of transverse channel velocity fluctuations. In a more systematic way, using the EFT ideas developed in 	\cite{Delacretaz:2021qqu}, it would be interesting to construct the Schwinger-Keldysh EFT associated with the BDNK theory. See 		\cite{Jain:2023obu,Mullins:2023ott} for related recent works. A more general treatment requires including fluctuations of the temperature as well as the longitudinal component of the fluid velocity. We hope we turn to this issue in the future. 
	
	\section*{Acknowledgment}
	\label{}
	We thank Lorenzo Gavassino,  Matthias Kaminski, and Omid Tavakol for valuable discussions and comments.  We especially thank Pavel Kovtun for discussions on ref.~\cite{Kovtun:2019hdm} and Hofie S. Hannesdottir for discussing ref.~\cite{Hannesdottir:2022xki}.
	NA was supported by grant number 561119208 ``Double First Class'' start-up funding of Lanzhou University, China. ST was supported by
	Younger Scientist Scholarship  QN2021043004 with the funding number E11I631KR0.

	\appendix
	\section{Equilibrium Fluctuations}
	\label{hydro_fluc_App}
	The problem of finding the mean value of quantities in equilibrium can be studied via finding the probability distribution of their deviations from the associated mean values \cite{Landau_1}. In the simplest setting,  quantities of interest are energy density and momentum density.
	What we need then is to exploit a fundamental property of entropy: probability of finding a  subsystem at energy density $\mathcal{E}$ and momentum density  $\mathcal{P}$ is given in terms of the total entropy of the system together with the medium, $S_t$, and also with its change in fluctuations:
	\begin{equation}\label{w_distribution}
		\text{w}(\mathcal{E},\mathcal{P}_i)\sim e^{S_t}\sim e^{\Delta S_t}  \sim e^{-\int_{\text{x}}\mathcal{R}/T} 
	\end{equation}
	where $\mathcal{R}=\Delta \mathcal{E}-\textbf{v}\cdot\Delta\boldsymbol{\mathcal{P}}-T \Delta \mathcal{S}$, with $\textbf{v}$ and $T$ being the equilibrium (mean) values of the velocity and the temperature of the system. $\Delta \mathcal{E}$, $\Delta\boldsymbol{\mathcal{P}}$, and $\Delta \mathcal{S}$ are the fluctuations of energy density, momentum density and the entropy density. Expanding the latter to second order in terms of the former two ones, and using the fact that first derivatives vanish in equilibrium, we find
	\begin{equation}\label{R/T}
		\frac{\mathcal{R}}{T}\sim \frac{1}{2}\bigg[-\delta\left(\frac{1}{T}\right)\Delta\mathcal{E}+\delta\left(\frac{\textbf{v}}{T}\right)\cdot\Delta\boldsymbol{\mathcal{P}}\bigg]
	\end{equation}
	where $\delta$ and $\Delta$ indicate the fluctuation in source and the fluctuation in energy/momentum density, respectively.
	When $\textbf{v}=\textbf{0}$ in equilibrium, $\Delta\mathcal{E}=c_V \delta T$ and $\Delta\boldsymbol{\mathcal{P}}=w \delta\textbf{v}$. Therefore \eqref{R/T} becomes
	\begin{equation}\label{}
		\frac{\mathcal{R}}{T}\sim \frac{1}{2}\bigg[\frac{c_V}{T^2}(\delta T)^2+\frac{w}{T}(\delta\textbf{v})^2\bigg]
	\end{equation}
	Substituting this into \eqref{w_distribution}, 
	the ``equal-time" correlation functions can be read immediately as follows \cite{Landau_2}:
	\begin{subequations}\label{}
		\begin{align}\label{Correlation_Equ_vv}
			\langle \delta v_i(\textbf{x}_1)\delta v_j(\textbf{x}_2)\rangle=&\,\delta_{ij}\frac{T}{w}\,\delta^3 (\textbf{x}_1-\textbf{x}_2)\,,\\ \label{Correlation_Equ_TT}
			\langle \delta T(\textbf{x}_1)\delta T(\textbf{x}_2)\rangle=&\,\frac{T^2}{c_V}\,\delta^3 (\textbf{x}_1-\textbf{x}_2)\,.
		\end{align}
	\end{subequations}

\section{Hydrodynamic Fluctuations}
\label{}
Now let us move on to study the ``different-time" correlation functions $\langle \delta \phi_{a}(t, \textbf{x}) \delta \phi_{b}(0, \textbf{0})\rangle $.
	It is easy to show that 
	\begin{equation}\label{one_sided}
		\big\langle\delta \phi_{a}\delta \phi_{b}\big\rangle_{\omega\textbf{k}}=\big\langle\delta \phi_{a}\delta \phi_{b}\big\rangle^{(+)}_{\omega\textbf{k}}+\big\langle\delta \phi_{b}\delta \phi_{a}\big\rangle^{(+)}_{-\omega \,-\textbf{k}}
	\end{equation}
	where $\big\langle\delta \phi_{a}\delta \phi_{b}\big\rangle^{(+)}_{\omega\textbf{k}}$ is one-sided Fourier transformation of $\langle \delta \phi_{a}(t, \textbf{x}) \delta \phi_{b}(0, \textbf{0})\rangle $, defined as \eqref{Fourier}, but with the time integral taken from $0$ to $+\infty$ \cite{Landau_1,Landau_2}.
	The strategy is to find equations between one-sided Fourier transforms and then to calculate correlation functions by use of \eqref{one_sided}. 
	
		We also take the momentum to be directed in the $z-$direction $\qn=(0,0,q)$. This then allows us to calculate the correlation functions in two channels. In the longitudinal (sound) channel, the correlation functions of  $\delta T/T$ and $\delta v^{\parallel}\equiv\delta v_{z}$ are involved. In the transverse (shear) channel, only $\langle \delta v^{\perp} \delta v^{\perp} \rangle$ needs to be computed.
	\subsection*{Longitudinal channel}
	Multiplying the two equations \eqref{EoM_t} and the $z-$component of \eqref{EoM_z} by $\delta T(0,\textbf{0})$ from the right side, and averaging over the local equilibrium, we find
	the equations of one-sided Fourier transformations as the following:
	\begin{equation}\label{Long_Channel_eq_set_1}
		\begin{split}
			\left[-i \wn-\frac{1}{12}(\theta_{\eta}\qn^2+3\pi_{\eta}\wn^2)\right]\langle
			\frac{\delta T}{T}\frac{\delta T}{T}\rangle^{(+)}+
			\bigg[\frac{i q}{3}+\frac{1}{12}(\theta_{\eta}+\pi_{\eta})q\wn\bigg]&\langle
			\delta v_z \frac{\delta T}{T} \rangle^{(+)}=\frac{\bar{\eta}(4-i\pi_{\eta}\wn)}{48\bar{p}^2\,T^4}\\
			\bigg[- i \wn+\frac{1}{3}\qn^2-\frac{1}{12}(\pi_{\eta}\qn^2+3 \theta_{\eta}\wn^2)\bigg]\langle
			\delta v_z\frac{\delta T}{T}\rangle^{(+)}+
			\bigg[i q+\frac{1}{4}(\theta_{\eta}+\pi_{\eta})q\wn\bigg]&\langle
			\frac{\delta T}{T}\frac{\delta T}{T}\rangle^{(+)}=\,0
		\end{split}
	\end{equation}
	Note that the right side of the first equation above is the consequence of using \eqref{Correlation_Equ_TT}. 
	When multiplying \eqref{EoM_t} and the $z-$component of \eqref{EoM_z}  by $\delta v_z(0,\textbf{0})$, and repeating the above process, we arrive at
	\begin{equation}\label{Long_Channel_eq_set_2}
		\begin{split}
			\left[-i \wn-\frac{1}{12}(\theta_{\eta}\qn^2+3\pi_{\eta}\wn^2)\right]\langle
			\frac{\delta T}{T}\delta v_z\rangle^{(+)}+\left[\frac{i q}{3}+\frac{1}{12}(\theta_{\eta}+\pi_{\eta})q\wn \right]&\langle
			\delta v_z\delta v_z\rangle^{(+)}=0\\
			\bigg[- i \wn+\frac{1}{3}\qn^2-\frac{1}{12}(\pi_{\eta}\qn^2+3 \theta_{\eta}\wn^2)\bigg]\langle
			\delta v_z\delta v_z\rangle^{(+)}+
			\bigg[i q+\frac{1}{4}(\theta_{\eta}+\pi_{\eta})q\wn\bigg]&\langle
			\frac{\delta T}{T}\delta v_z\rangle^{(+)}=\,\frac{\bar{\eta}(4-i\theta_{\eta}\wn)}{16\bar{p}^2\,T^4}
		\end{split}
	\end{equation}
	From these equations we find
	\begin{subequations}\label{}
		\begin{align}\label{delta_delta}
			\langle
			\frac{\delta T}{T}\frac{\delta T}{T}\rangle^{(+)}=&\,\frac{\bar{\eta}}{4\bar{p}^2T^4}\frac{1}{\mathcal{D}_L(\wn,\qn)} \,(i\pi_{\eta} \wn-4)\,\big((\pi_{\eta}-4)\qn^2+12i \wn+3 \theta_{\eta}\wn^2\big)\\\label{}
			\langle\delta v_z\frac{\delta T}{T}\rangle^{(+)}=&\,\frac{3\bar{\eta}}{4\bar{p}^2T^4}\frac{1}{\mathcal{D}_L(\wn,\qn)} \,iq\,(i\pi_{\eta} \wn-4)\,\big(4-i (\theta_{\eta}+\pi_{\eta})\wn\big)\\\label{}\langle
			\frac{\delta T}{T}\delta v_z\rangle^{(+)}=&\,\frac{3\bar{\eta}}{4\bar{p}^2T^4}\frac{1}{\mathcal{D}_L(\wn,\qn)} \,iq\,(i\theta_{\eta} \wn-4)\,\big(4-i (\theta_{\eta}+\pi_{\eta})\wn\big)\\\label{}\langle
			\delta v_z\delta v_z\rangle^{(+)}=&\,\frac{3\bar{\eta}}{4\bar{p}^2T^4}\frac{1}{\mathcal{D}_L(\wn,\qn)} \,(i\theta_{\eta} \wn-4)\,\big(\theta_{\eta}\qn^2+12i \wn+3 \pi_{\eta}\wn^2\big)
		\end{align}
	\end{subequations}
	with 
	\begin{equation}\label{Denominator_energy}
		\begin{split}
			\mathcal{D}_L(\wn, \qn)=&\,9 \pi _{\eta }  \,\theta _{\eta }\,\wn^4+36i  \left(\theta _{\eta }+\pi _{\eta }\right)\, \wn^3-6 \left(\pi _{\eta } \left(\theta _{\eta }+2\right) \qn^2+24\right)\wn^2\\
			&-12 i   \left(\theta _{\eta }+\pi _{\eta }+4\right) \qn^2\wn+ \left(\left(\pi _{\eta }-4\right) \theta _{\eta } \qn^2+48\right)\qn^2
		\end{split}
	\end{equation}
	Then it is easy to calculate the correlation functions by using \eqref{one_sided}. We do not represent those expressions here explicitly. 
	\subsection*{Transverse channel}
	\label{App_trans}
	The correlation functions in the transverse channel are not complicated, even at $\textbf{v}_0\ne 0$. Therefore, we continue to express the result in general $\textbf{v}_0$.
	
	Due to the rotational symmetry in the transverse plane,  we only need to consider the $x-$component of \eqref{EoM_z}. Multiplying by $\delta v_x(0,\textbf{0})$, the one-sided Fourier transformation of $\langle \delta v_{x}(t,\textbf{x})\delta v_{x}(0,\textbf{0})\rangle$ is found to be  
	
	\begin{equation}\label{delta_v_delta_v_plus}
		\langle\delta v_i\delta v_i\rangle^{(+)}=\,\frac{i\bar{\eta}}{4\bar{p}^2T^4}\frac{4i \gamma+\wn+\gamma^2(\theta_{\eta} -1)\wn}{\mathcal{D}_T(\wn,\qn)}\,\,\,\,\,\,i=x,y
	\end{equation}
	where $\mathcal{D}_T$ is given in the main text.
	Again, it is easy to calculate $\langle\delta v_i\delta v_i\rangle$ by using \eqref{one_sided}.
	
	\section{Landau equations}
	\label{Landau_App}
In this appendix we briefly review how to use the Landau equations to find the singularities of a Feynman diagram.
	\vspace{3mm}
	\subsection{threshold singularities}
	\label{}

	Let us suppose an arbitrary Feynman diagram (in the scalar theory) is represented by the following integral
	\begin{equation}\label{Landau_int}
	\mathcal{I}(p)=\,	\int\frac{d^4k}{(2\pi)^4}\frac{d^4l}{(2\pi)^4}\cdots\frac{B}{A_1\,A_2\,\cdots}
	\end{equation}
	where $A_i=q_i^2+m^2$ and $q_i$ is the four-momentum corresponding to the given line in the diagram. In addition, $B$ is a polynomial of the four-vectors $q_i$. Using the well known ``Feynman' parameters'' method, one may write
	\begin{equation}\label{alpha_int}
		\frac{1}{A_1A_2\cdots A_n}=(n-1)!\int_{0}^{1}\int_{0}^{1}\cdots\int_{0}^{1}\frac{d\alpha_1d\alpha_2\cdots d\alpha_n{\delta(\alpha_1+\alpha_2+\cdots+\alpha_n-1)}}{\big(\alpha_1 A_1+\alpha_2 A_2+\cdots \alpha_n A_n\big)^n}
	\end{equation}
	Landau argues that the singularities of the integral \eqref{Landau_int}\footnote{These singularities include both normal (discussed in this paper) and anomalous  thresholds. For anomalous thresholds  see \cite{Nambu:1958,Karplus:1958zz}.} are actually a solution to the following equations/conditions \cite{Landau:1959} :
	\begin{enumerate}
		\item For each propagator $i=1,\cdots,n$:
		\begin{equation}\label{Landau_on_shell}
			q_i^2+m^2=0 \,\,\,\,\,\,\,	\text{or}  \,\,\,\,\,\,\,\,   \alpha_i=0\,.
		\end{equation}
		\item For each loop momentum integration variable $k_j$: 
		\begin{equation}\label{Landau_loop}
			\frac{\partial}{\partial k_j}\sum_{i} \alpha_i (q_i^2+m^2)=0.
		\end{equation}
	\end{enumerate}
	These conditions are referred to as the ``\textit{Landau equations}'' or ``\textit{Landau conditions}''.  The first condition,  \eqref{Landau_on_shell}, is simply the on-shell condition for the internal momenta.  	These on-shell conditions identify where singularities can occur in the integrand in a Feynman integral, which is a necessary but not a sufficient condition for a singularity to develop in the full integral \cite{Hannesdottir:2022xki}.
	 The second one, given by \eqref{Landau_loop},  is called the ``Landau loop equation". We explain it below.
	\vspace{2mm}

	To understand \eqref{Landau_loop}, let us denote that the singularity of the integrand will be the singularity of the integral if it cannot be avoided by doing any contour deformation. One way to achieve this is that, by changing the external momentum, the singularity of the integrand reaches the endpoint of the integration;  the latter is called ``\textit{end-point singularity}'' which is actually quite easy to identify. Another possibility is that, for a given set of external momenta,  the contour of integration is pinched between singularities of the integrand at some real value of the  loop momentum.  Remember when two curves are tangent; their normal vectors at the point of intersection are parallel. Similarly,  there will be a contour pinch when the gradients of the on-shell conditions (with respect to the independent loop momenta) are linearly dependent. The latter singularity is called ``\textit{pinch singularity}''  \cite{Hannesdottir:2022xki,Eden:1966} \footnote{See also \cite{Windisch:2013dxa,Huber:2022nzs}.}.
		\vspace{2mm}
		
It should be noted that solutions to the Landau equations that require some Feynman
	parameters $\alpha_i$ to be either negative or complex do not correspond to singularities of $\mathcal{I}(p)$ that
	can be accessed with real on-shell \textit{external} momenta \cite{Hannesdottir:2022xki}. The latter momenta define the \textbf{physical region}. On the other hand, when $\mathcal{I}(p)$ is multivalued, one can be in the physical region on different Riemann sheets:
	\begin{itemize}
		\item The entire set of complex points accessible through the analytic continuation of the physical region defines the \textbf{physical sheet}. Singularities associated with positive Feynman parameters ($\alpha_i>0$) are on the physical sheet and are called \textbf{normal threshold singularities}.
		
		\item Singularities associated with negative or complex values of the Feynman parameters  are not on the physical sheet and are called \textbf{pseudo-threshold singularities.}
		\end{itemize}
	The above singularities are all called \textbf{first-type Landau singularities}. As we saw, for this class of singularities the notion of  threshold in the space of external variables is well-defined. The Landau equations however can also describe singularities at infinite loop momenta. Remember the fourth comment below \eqref{S_qn_aplplied}. We saw that the two singularities $\wn_{12}$ and $\wn_{21}$ were of this type. Sometimes they are referred to as \textbf{second-type singularities} \cite{Eden:1966}.
	\subsection{Nature of singularities}
	\label{}
	Let us combine \eqref{Landau_int} and \eqref{Landau_int}:
	\begin{equation}\label{}
		\mathcal{I}(p)=\,(n-1)!	\int\frac{d^dk}{(2\pi)^d}\frac{d^dl}{(2\pi)^d}\cdots\int_{0}^{1}\int_{0}^{1}\cdots\int_{0}^{1}\frac{d\alpha_1\cdots d\alpha_{n}{\delta(\alpha_1+\cdots+\alpha_{n}-1)}}{f^n}
	\end{equation}
	where $f=\alpha_1 A_1+\cdots \alpha_{n} A_{n}$. Let us call the solutions to the Landau-loop equations $\alpha^*$. Landau argues that leading singularity of $	\mathcal{I}(p)$ is given by 
	\begin{equation}\label{}
\sim\big( f(p,\alpha^*)\big)^{\frac{1}{2}m-n}
	\end{equation}
	where $m$ is the number of independent integrations in  $	\mathcal{I}(p)$ and $n$ is the number of internal lines in Feynman diagram.
	\begin{itemize}
		\item The number of momentum integration in $d-$dimension is $d\nu$ where $\nu$ is the number of independent contour. Clearly, $\nu=n-V+1$ with $V$ being the number of vertices in the diagram. Thus the number of momentum integration in $d-$dimension is $d(n-V+1)$.
		\item There are $n-1$ of $\alpha$-integrations.
		\end{itemize}   
	Therefore, $m$, the total number of independent intgartions is given by:  
	\begin{equation}\label{}
m= d(n-V+1)+ (n-1)\,.
	\end{equation}
		Instead of number of contours, it is more convenient to represent it  in terms of the number of vertices $V$. It then reads
	\begin{equation}\label{}
		\sim\big( f(p,\alpha^*)\big)^{\frac{d-1}{2}(n+1)-\frac{d}{2}V}\,.
	\end{equation}

	\providecommand{\href}[2]{#2}\begingroup\raggedright\endgroup
	
\end{document}